\documentclass{pramana}
\usepackage[linesnumbered,ruled,vlined]{algorithm2e}%
\usepackage{graphicx,amsmath,bm}
\usepackage{cite}
\usepackage{tikz}
\usetikzlibrary{quantikz}
\usepackage{tikz-cd}
\newsavebox{\boxA}
\newsavebox{\boxB}
\newsavebox{\boxC}
\newsavebox{\boxD}
\newsavebox{\boxE}
\newsavebox{\boxF}
\newsavebox{\boxG}
\newsavebox{\boxH}
\begin{document}\sloppy

%%paper title
%%For line breaks \\ can be used within title 
\title{Quantum Computation of Fluid Dynamics}

%%author names are separated by comma (,) 
%%use \and before the last author name 
%%\textsuperscript{number} is used for affiliation
%%use a * along with the number separated by comma
%% for the  author for correspondence

\author{{Sachin S. Bharadwaj}\textsuperscript{1}\and  {Katepalli R. Sreenivasan}\textsuperscript{1,2,3,4*} }
\affilOne{\textsuperscript{1} Department of Mechanical and Aerospace Engineering, New York University, New York 11201 USA\\}
\affilTwo{\textsuperscript{2} Department of Physics, New York University, New York 10012 USA\\}
\affilThree{\textsuperscript{3} Courant Institute of Mathematical Sciences, New York University, New York 10012 USA\\}
\affilFour{\textsuperscript{4} New York University, Abu Dhabi 129188, UAE\\}
%, New York University New York NY 11201 USA
%%escape two column mode for title, affiliation and abstract
%%by giving \twocolumn command as shown

\twocolumn[{

\maketitle

%%include \corres to print the corresponding author Email id
\corres{katepalli.sreenivasan@nyu.edu}

%%include \msinfo for
%%manuscript information such as
%%received, revised and accepted dates
%%
\msinfo{15 June 2020}{15 June 2020}{15 June 2020}

%%abstract
\begin{abstract}
Studies of strongly nonlinear dynamical systems such as turbulent flows call for superior computational prowess. With the advent of quantum computing, a plethora of quantum algorithms have demonstrated, both theoretically and experimentally, more powerful computational possibilities than their classical counterparts. Starting with a brief introduction to quantum computing, we will distill a few key tools and algorithms from the huge spectrum of methods available, and evaluate possible approaches of quantum computing in fluid dynamics.

%Studying non-linear dynamical systems, for instance, problems that deal with turbulent flow physics, calls for superior means of computational prowess. With the onset of quantum computing, a plethora of quantum algorithms have demonstrated both theoretically and experimentally to have higher computational efficiency than their classical counterparts. Here, starting with a brief introduction to the field of quantum computing, we distill out, from the huge spectrum of quantum computational methods, some of the key computational tools and algorithms and also enlisting and evaluating possible approaches that might be amenable for beginning a new era of quantum computing fluid dynamics. 
\end{abstract}

%%insert keywords separated by comma using \keywords{words}

%%include \pacs{number} to print the PACS number

}]
%%close the twocolumn escape here

%%include \doinum{number}for the DOI number in the header
%%include \volnum{number} for the volume number in the header
%%include \year{yyyy} for  year of publication in the header
%%include \pgrange{num--num} page range of article in the header
%%include \artcitid{num} for the article citation id
%%include \lp to print last page of the article
%%include \setcounter{page}{pagenum} for the exact starting page of the article

\doinum{12.3456/s78910-011-012-3\\
\textit{This is a pre-print of an article to appear in Springer-Pramana-Journal of Physics}}
\artcitid{\#\#\#\#}
\volnum{123}
\Year{2020}
\pgrange{1--20}
\setcounter{page}{1}
\lp{20}

\section{INTRODUCTION}

Fluid mechanics as a field poses a vast array of interesting questions that relate to almost everything we see around us. Apart from theory and experiments, computational methods have greatly aided fluid mechanics research over the past few decades; indeed, with the growth in computers of increasingly higher computational power, fluid mechanical simulations have become highly realistic. But with increasing sophistication comes new generations of questions. For instance, even with the great advances seen in High Performance Computing (HPC), and despite the progress being made continually by very large Direct Numerical Simulations (DNS), one cannot say that long standing questions relating to the separation of scales in turbulence have been addressed fully. Without necessarily making the explicit case that computer technology development has hit obstacles, we simply note that the computational challenges being faced at present are so enormous that simply making supercomputers more powerful cannot catch up with the demands. Not only manufacturing smaller transistors face quantum effects, but also their integration into massively complex systems poses numerous challenges. To break this barrier, one needs a change of paradigm in computing. Enter quantum computing! 

In quantum computing, we manipulate quantum systems to perform calculations and simulations. We are thus entering an era in which computations are becoming more ``physical". In fact, it was a dream of Richard Feynman \cite{feynman1999simulating} to simulate a quantum system by using another quantum system. We are now in the NISQ (Noisy Intermediate Scale Quantum) era \cite{preskill2018quantum}, where we have quantum computers of sizes ranging from 50 qubits to a few hundreds of them. (Qubits are essentially the quantum analogue of classical bits and will be described later; it suffices to say here that their number characterizes the power and size of a quantum computer.) The word ``noisy" indicates that quantum devices are still prone to errors from external and internal noises, and are not yet perfect. Yet, with quantum devices of the size just emerging, quantum computing (QC) can outperform many operations that current supercomputers strain to achieve. Quantum Computers have already started demonstrating their practicability in various fields such as finance strategies, medicine, quantum materials and chemical simulations, resource management, optimization and cryptography. What we wish to investigate here is its utility for performing Computational Fluid Dynamics (CFD) research.

This paper presents an outlook on doing CFD quantum mechanically, which we term Quantum Computation of Fluid Dynamics (QCFD). It introduces and motivates researchers who wish to study fluid mechanics or dynamical systems, in general, to the new possibility of using quantum computers. In section 2, we present a brief overview of QC and its differences from classical computing. We then set up in section 3 the big picture of how fluid mechanics study can be viewed in the QC context. This is followed by a description of methods that are lattice based (section 4) and continuum based (section 5). Section 5 also touches on the possibility of studying quantum turbulence and reviews existent methods and proposes newer directions. In section 6, we list from the horde of QC algorithms a few key ones that are deemed important for our purposes, and provide a few specific demonstrations. Finally, we briefly mention in section 7 the currently available quantum machines and quantum programming, ending with a few conclusions on QCFD in section 7.

\section{OVERVIEW OF QUANTUM COMPUTATION}

The purpose of this section is to provide a brief overview of how the working rules of QC differ from those of classical computing. It is intended for readers with minimal background in quantum computing; a detailed account can be found in \cite{nielsen2002quantum}. 

\subsection{What is Quantum Computation?} It is a form of computation centered on quantum mechanics, manipulating information in the form of quantum bits called ``qubits", by designing appropriate ``quantum algorithms" that comprise ``quantum gates and circuits", which in turn act on these qubits to yield the intended result. This sounds similar to classical computation, except that every word or phrase is prefixed by the word ``quantum". We shall explain each of these terms below. 
\begin{figure}[htpb!]
\centering{
\includegraphics[scale=0.4]{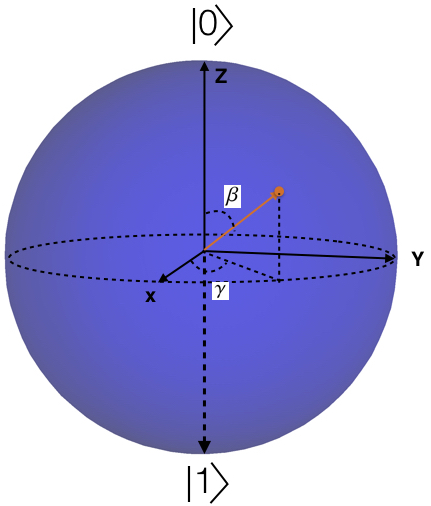}
}
\caption{Bloch Sphere. Here $\alpha$ is the global phase, $\beta$ is the polar angle and the azimuthal angle $\gamma$ is the relative phase}
\end{figure}
\subsection{Qubits} Qubits form the work horse of quantum computation. Similar to classical bits, quantum bits are objects that hold information describing quantum physical systems, which are eventually manipulated to perform a computational task. In reality, these qubits represent the state of an actual quantum physical system, governed by laws of quantum mechanics. Mathematically, it is given by the wavefunction $\Psi$, which completely encodes all the details describing the state of a quantum object. As a working example, a qubit could represent the two spin states of an electron. There exist several physical realizations of qubits such as Quantum Electro-Dynamic (QED) Optical Cavities, Ultra Cold atoms and Rydberg ions, Superconductors and Topological Materials (Majorana fermions), Photons, Quantum dots, Nuclear Magnetic Resonance (NMR), etc. For the rest of the paper, however, we shall simply describe qubits as abstract mathematical objects. For now and for all practical purposes, we shall denote qubits as wavefunctions, which are vectors in a complex vector space called the Hilbert Space $\mathbb{H}$. In Dirac's bra-ket notation, it is represented as a ``ket" vector $|\Psi\rangle$ in $\mathbb{H}$ $(\in \mathbb{C}^{n})$ 
\begin{equation}
    |\psi\rangle = \begin{bmatrix}
           c_{1} \\
           c_{2} \\
           \vdots \\
           c_{n}
         \end{bmatrix}; c_{i} \in \mathbb{C}
\end{equation}
while ``bra", given by $\langle\Psi|$, is the vector dual $= |\Psi\rangle^{\dag}$. These wavefunctions obey all the rules of a complex vector space. An obvious extension to this concept is to multiple qubits by taking tensor products of individual wavefunctions, which together lie in a tensor-product Hilbert space of corresponding wavefunctions: $|\Psi\rangle = (|\psi_{1}\rangle \otimes ... \otimes |\psi_{n}\rangle) \in \mathbb{H}^{\otimes n}$. For instance, the two spin states (spin-up $\uparrow$ and spin-down $\downarrow$) of an electron, could correspond to the state eigenvectors $|0\rangle$ and $|1\rangle$, respectively. The wavefunction of such two level or two state systems is a complex vector in $\mathbb{H} \in \mathbb{C}^{\otimes 2} $ which, when expressed mathematically as a linear combination of the basis vectors, has the form
\begin{equation}
    |\Psi\rangle = c_{1}|0\rangle + c_{2}|1\rangle,
\end{equation}
where 
\begin{equation}
|0\rangle = \begin{bmatrix} 1  \\ 0\end{bmatrix} \text{ and }
|1\rangle = \begin{bmatrix} 0  \\ 1 \end{bmatrix}.
\end{equation}
\begin{figure}[htpb!]
\centering{
\includegraphics[scale=0.95]{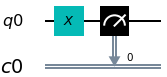}
}
\caption{Circuit for a simple NOT operation}
\end{figure}
\begin{figure}[htpb!]
\centering{
\includegraphics[scale=0.5]{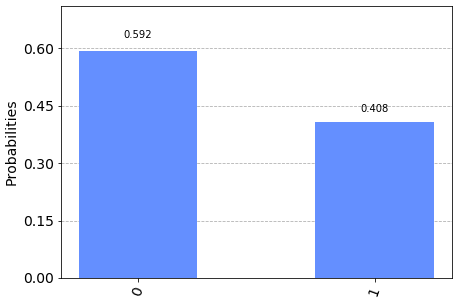}
}
\caption{Probabilities before NOT operation }

\end{figure}
\begin{figure}[htpb!]
\centering{
\includegraphics[scale=0.5]{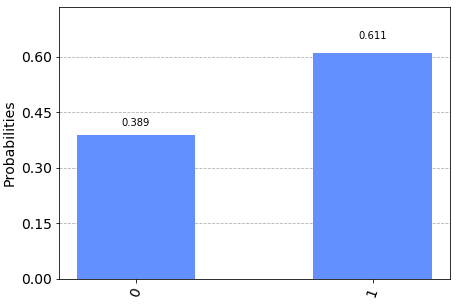}
}
\caption{Probabilities after NOT operation }
\end{figure}

Physically, the complex numbers $c_{1}$ and $c_{2}$ represent the \textit{probability amplitudes} of the electron being in a given basis state, whose squares, $|c_{1}|^2$ and $|c_{2}|^2$, according to Born's principle, give the \textit{probability} of the electron being in either state, $|0\rangle$ or $|1\rangle$. This only implies, at any given time, that the electron has a finite probability of being in both states simultaneously, unlike an unbiased coin. (Such superposition states sum to unity as probabilities should.) This is what distinguishes a classical state from a quantum one: a quantum state prior to measurement or observation can exist in a superposition of two different states, while a classical state can be in only one of them at a given time. This gives us access to multiple basis states simultaneously; this is the quantum parallelism which we shall examine subsequently. Apart from the Hilbert space representation, another useful visualisation of a qubit is the Bloch Sphere representation shown in Figure 1. Every possible state described by equation (2), can be represented as a vector on this unit sphere via the relation
\begin{equation}
    |\Psi\rangle = e^{i\alpha}\Big[\cos\Big(\frac{\beta}{2}\Big)|0\rangle + e^{i\gamma}\sin\Big(\frac{\beta}{2}\Big)|1\rangle \Big],
\end{equation}
where $\alpha$ is the global phase, $\gamma$ the relative phase (azimuthal angle) and $\beta$ the polar angle. All transformations and actions of ``quantum gates" on qubits are to be regarded as rotational affine transformations on the Bloch vector. 

%\begin{figure*}
%\centering\includegraphics[height=11cm,width=16cm]{PNLD_gates.jpeg}
%\caption{Basic Quantum Logic Gates }
%\end{figure*}
\savebox{\boxA}{\begin{quantikz} {\lstick{}   & \gate{X} &  \qw}\end{quantikz}}%

\savebox{\boxB}{ \begin{quantikz} {\lstick{}   & \gate{Y} &  \qw}\end{quantikz}}%

\savebox{\boxC}{ \begin{quantikz} { \lstick{}   & \gate{Z} &  \qw}\end{quantikz}}%

\savebox{\boxD}{ \begin{quantikz}{ \lstick{}   & \gate{H} &  \qw}\end{quantikz}}%

\savebox{\boxE}{\begin{quantikz} {\lstick{}   & \gate{T} &  \qw}\end{quantikz}}%

\savebox{\boxF}{ \begin{quantikz} \lstick{\ket{q\textsubscript{1}}}& \ctrl{1} & \qw \\
\lstick{\ket{q\textsubscript{2}}} & \targ{} & \qw 
 \qw\end{quantikz}}%

\savebox{\boxG}{ \begin{quantikz} \lstick{\ket{q\textsubscript{1}}}& \swap{1}  & \qw \\
\lstick{\ket{q\textsubscript{2}}}& \targX{} & \qw 
 \qw\end{quantikz} }%

\savebox{\boxH}{ \begin{quantikz} \lstick{\ket{q\textsubscript{1}}}& \ctrl{1} & \qw \\
\lstick{\ket{q\textsubscript{2}}}& \ctrl{1} & \qw  \\
\lstick{\ket{q\textsubscript{3}}} & \targ{} & \qw
 \qw\end{quantikz}}%

\begin{table*}[htpb!]
\centering
\caption{Quantum Logic Gates}
\vspace{4mm}
\begin{tabular}{||c c c||} 
 \hline
 \textbf{Quantum Logic Gate} & \textbf{Circuit Symbol} & \textbf{Operation} \\ [0.5ex] 
 \hline\hline
 \textbf{X}(Pauli X) &\usebox\boxA & $|0\rangle \rightarrow |1\rangle$ \hspace{1cm}  $|1\rangle \rightarrow |0\rangle$ \\ 
 \hline
 \textbf{Y}(Pauli Y) & \usebox\boxB & $|0\rangle \rightarrow i|1\rangle$ \hspace{0.9cm}  $|1\rangle \rightarrow -i|0\rangle$  \\
 \hline
 \textbf{Z}(Pauli Z) & \usebox\boxC & $|0\rangle \rightarrow |0\rangle$ \hspace{1cm}  $|1\rangle \rightarrow -|1\rangle$  \\
 \hline
 \textbf{H}(Hadamard) & \usebox\boxD & $|0\rangle \rightarrow \frac{|0\rangle + |1\rangle}{\sqrt{2}}$ \hspace{1cm} $ |1\rangle \rightarrow \frac{|0\rangle - |1\rangle}{\sqrt{2}}$ \\
 \hline
 $\mathbf{R}_{\phi}$(Phase Shift) & \usebox\boxE & $|0\rangle \rightarrow |0\rangle$ \hspace{1cm} $ |1\rangle \rightarrow e^{i\phi}|1\rangle$  \\ [1ex] 
\hline
\textbf{CNOT} & \usebox\boxF
  & $|q_{1},q_{2}\rangle \rightarrow |q_{1},q_{2} \oplus q_{1}\rangle$ \\ 
\hline
\textbf{SWAP} & \usebox\boxG  & $|q_{1},q_{2}\rangle \rightarrow |q_{2},q_{1}\rangle$ \\
\hline
\textbf{Toffoli} &\usebox\boxH & $|q_{1},q_{2},q_{3}\rangle \rightarrow |q_{1},q_{2},q_{3} \oplus q_{1}\cdot q_{2}\rangle$ \\
\hline
\end{tabular}
 \end{table*}
 
\subsection{Quantum Gates, Circuits and Algorithms} 
In analogy to classical computing, where we write algorithms to manipulate information, and accomplish them fundamentally via logic gates such as AND, NOT, OR, NAND and Toffoli gates, information manipulation is accomplished by quantum algorithms using quantum logic gates and circuits, as explained below.

Quantum gates are fundamentally unitary operators U ($\text{U}\text{U}^{\dag}=\mathbb{I}$), which cause affine rotational transformations on qubits. These unitaries are linear and reversible operations (unlike classical gates such as NOT), and are also norm-preserving. It is obvious that an infinitely many such unitary transforms can exist but, among them, the fundamental and important ones are listed in Table 1. As in classical computing there are a set of quantum gates which are universal, and a detailed description can be found in \cite{nielsen2002quantum}. 

Now, a combination of such gates forms a quantum circuit. For instance, consider the quantum version of the classical NOT gate acting on a qubit. This is the X gate, given by $\sigma_{x}$, the Pauli operator. To see this in action, let us prepare a single qubit (q0) in the state $|\psi\rangle = \sqrt{0.6} |0\rangle + \sqrt{0.4} |1\rangle$ and then apply the \newline X gate = $\sigma_{x}$ =
$\begin{bmatrix}0 & 1 \\ 1 & 0\end{bmatrix}$. This yields
\begin{equation}
    \begin{bmatrix}0 & 1 \\ 1 & 0\end{bmatrix} \begin{bmatrix}1  \\ 0 \end{bmatrix} = \begin{bmatrix}0  \\ 1 \end{bmatrix}  \text{ and }   \begin{bmatrix}0 & 1 \\ 1 & 0\end{bmatrix} \begin{bmatrix}0  \\ 1 \end{bmatrix} = \begin{bmatrix}1  \\ 0 \end{bmatrix}.
\end{equation} 
That is, this gate just flips the state $|0\rangle \rightarrow |1\rangle$ and $|1\rangle \rightarrow |0\rangle$. After operating this gate, we measure the probabilities ($|c_{1}|^2$ and $|c_{2}|^2$) associated with the basis states and store them in a classical register (c0), as shown in Figure 2. The horizontal lines or circuit wires represent the time evolution of a qubit, and the double lines represent a classical bit. The meter symbol represents a measurement operation in the computational basis, while the X symbol denotes the quantum NOT gate. 

This circuit is now run on IBMQ Qiskit quantum simulator platform; from the results shown in Figures 3 and 4, it is clear that the states have flipped. It is worth noting that the NOT operation is applied on both $|0\rangle$ and $|1\rangle$ simultaneously, i.e., we now have the new state $|\psi'\rangle = \sqrt{0.4} |0\rangle + \sqrt{0.6} |1\rangle$. We shall explore the associated notion of quantum parallelism further but note here simply that a quantum circuit is basically a construction of quantum gates and wires that together act on a given set of qubits and perform the desired transformation, while a quantum algorithm is a collection of linked quantum circuits that performs a computational task. We shall outline the important algorithms in the sections to follow.

\subsection{Quantum Parallelism}
With a simple block diagram, we shall briefly outline quantum parallelism and the subtle difference between quantum and classical computing.

\begin{figure}[htpb!]
\centering{
\includegraphics[scale=0.45]{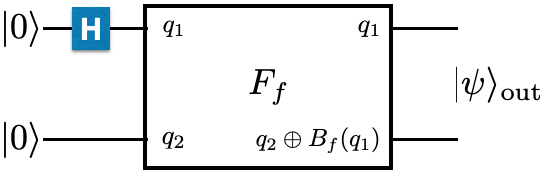}
}
\caption{Quantum Parallelism}
\end{figure}

Consider a Boolean function, B\textsubscript{f}: \{0,1\} $\mapsto$ \{0,1\}. We wish to evaluate this function for both 0 and 1 via quantum processing. For this, we take a 2-qubit state $|\psi_{in}\rangle=|q_{1}q_{2}\rangle$, where $q_{1},q_{2}\in\{0,1\}$. To keep our discussion brief, let us accept the existence of an oracle function (black-box) $F_{f}$ that basically performs $|q_{1,}q_{2}\rangle \xrightarrow{F_{f}} |q_{1},q_{2}\bigoplus\text{B}_{f}(q_{1})\rangle$, where $\bigoplus$ represents modulo 2 addition (refer to \cite{nielsen2002quantum} for the details of the black-box). To compute both B\textsubscript{f}(0) and B\textsubscript{f}(1) classically, we would have to do the computation twice. Now let us look at the quantum circuit in Figure 5, whose action is as follows: First, a Hadamard gate is applied on the first qubit: $|0\rangle\otimes|0\rangle\xrightarrow{H\otimes\mathbb{I}}\frac{|0\rangle+|1\rangle}{\sqrt{2}} \otimes|0\rangle$. This state now forms the input to the black-box, which finally produces the output state by applying $F_{f}$, as
\begin{equation}
    |\Psi\rangle_{out} = \frac{|0,\text{B}_{f}(0)\rangle+ |1,\text{B}_{f}(1)\rangle }{\sqrt{2}} .
\end{equation}
\begin{figure*}
\centering\includegraphics[scale=0.4]{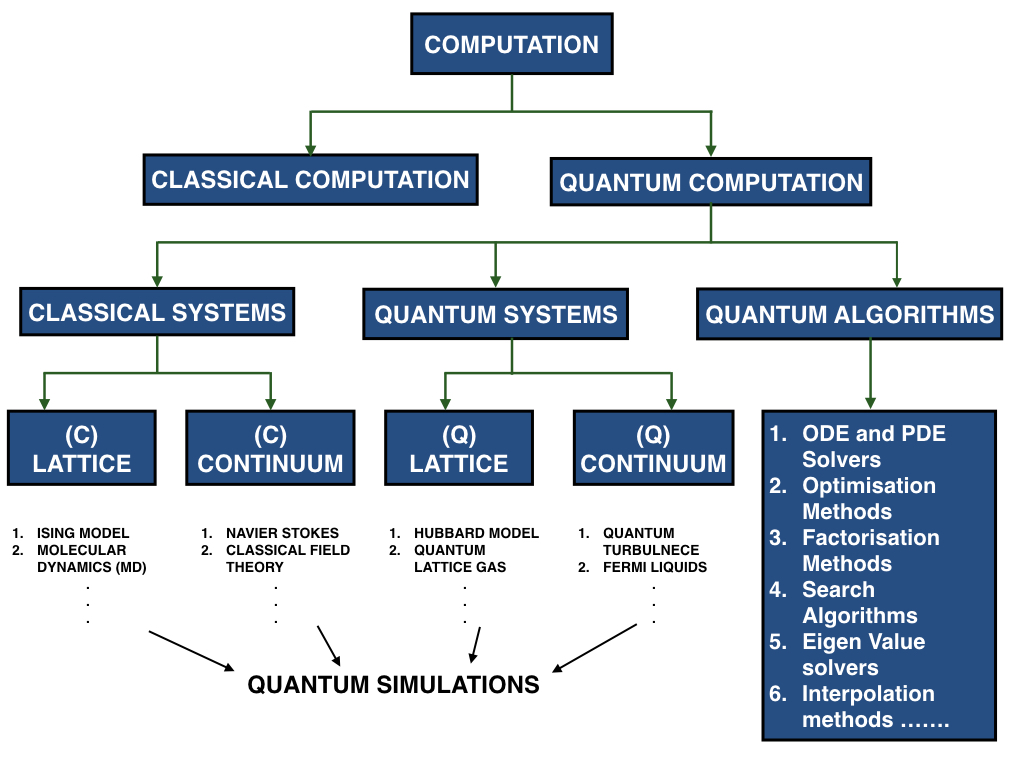}
\caption{Classification of problems based on QC methods}
\end{figure*}

Now, we have evaluated in just one shot both B\textsubscript{f}(0) and B\textsubscript{f}(1). This is quantum parallelism. The concept can be extended to more qubits, and also be used to extract information about some global properties of the function B\textsubscript{f}, so as to verify whether or not a given function is a constant; algorithms such as Deustch-Josza and Simon's algorithm can do that and more \cite{nielsen2002quantum}. This inherent parallelization of QC at the physical level demonstrates one of the many subtle and fundamental differences that sets QC apart from classical computations. In the sections to follow, we shall look at algorithms, gradually narrowing our scope to a discussion of fluid mechanics.

\section{QUANTUM COMPUTING OF DYNAMICAL SYSTEMS: THE BIG PICTURE}
We now attempt to address the task at hand: Analyzing the possible utility and advantages of quantum computing to study physical systems, fluid mechanical in particular. To this end, a slight digression towards a broader picture of QC study, shown in Figure 6, is useful for classifying the problems and methods for fluid mechanics. There are primarily three possible sets of problems which could be addressed by QC: (1) quantum systems, (2) classical systems and (3) quantum algorithms. Each of them is described below. 

\subsection{Quantum Systems}
Quantum systems are obvious candidates for using QC. Though all quantum systems are legitimate candidates, problems that are currently being explored, or could be explored, fall in two categories: \newline

\noindent(a) \textit{Lattice based systems}: Most hard quantum condensed matter systems such as the Hubbard problem or the quantum lattice gas fall in this category. Here, one can look at the lattice based Hamiltonians to either perform a quantum simulation or compute observables and properties via specific algorithms. \\

\noindent(b) \textit{Continuum problems}: On the other hand, some problems such as quantum turbulence and quantum liquids, would require the integration of the many-body Schr\"odinger equation followed by a mapping to macroscopic observables. One could also use quantum algorithmic numerical tools to integrate model equations such as Gross Pitaevski equations (in the case of quantum turbulence) or do a quantum Monte Carlo study, etc. 

\subsection{Classical Systems}
A slightly harder but an interesting avenue would be to compute classical systems using quantum computing. Most of the effort here would be spent in translating classical dynamics into the quantum language. One can then harness the quantum advantage from here on. Once again, a similar classification could be done, where one looks at lattice systems, Lattice Boltzmann Methods and molecular dynamics, or one can start using quantum based mathematical tools such as ODE solvers and eigenvalue solvers or optimization methods for integrating and solving classical governing equations such as the Navier-Stokes equations. 

\subsection{Computational Tools}
Though the development of quantum algorithms would need new mathematical tools, this step can proceed independently up to a certain point. Here one would be interested primarily in developing, quantum mechanically, the numerical solvers or methods available on classical machines, such as optimization, ODE/PDE solvers, factorizations, data search, eigenvalue solvers, etc. 

With this background, we now proceed to examine each of these methods and provide real quantum computational demonstrations. From these methods, we shall focus primarily on two methods suitable for studying fluid dynamics: (1) lattice based methods, and (2) continuum quantum simulations and quantum algorithms.

\section{LATTICE SIMULATIONS}
In addition to the popular computational methods such as (DNS) \cite{orszag1972numerical,rogallo1981numerical,pope2001turbulent,yeung2005high,iyer2020classical}, Large Eddy Simulations (LES) \cite{smagorinsky1993large,meneveau2000scale,pope2001turbulent}, Reynolds Averaged Navier-Stokes (RANS) \cite{pope2001turbulent,davidson2015turbulence} and other modelling techniques, the Lattice Boltzmann Method (LBM) \cite{succi1991lattice,succi2001lattice} has also been used recently to model fluid dynamical problems. The underlying principle governing LBM stems from the classical Boltzmann kinetic transport mechanism, which models the fluid as an ensemble of a large number of fictitious ``fluid particles" placed on a uniform lattice. These fluid particles advect in some allowed velocity directions and collide with each other resulting in a scattering-relaxation type process, which results to a net momentum transfer, as shown in Figure 7. The main advantage of this model is the large reduction in the number of degrees of freedom with which one would otherwise have to deal in the continuum case. The basic LBM equation is
\begin{equation}
   \underbrace{ \big(\partial_{t} + \mathbf{v_{\alpha} . \nabla}\big) \rho_{\alpha}(\mathbf{r},t)}_\text{Advection} =  \underbrace{S_{\alpha \beta} \big( \rho_{\beta}^{eq}(\mathbf{r},t) - \rho_{\beta}(\mathbf{r},t)\big)}_\text{Scattering-Relaxation} .
\end{equation}
where $\mathbf{v}$ is the velocity, $\rho$ is the mass density, $S$ is the scattering matrix and $\rho^{eq}$ is the equilibrium distribution of the mass density (for a detailed review refer to \cite{succi2001lattice}). One could naively say that, since quantum mechanical problems inherently deal with quantized ``particles", problems that involve particle tracking (e.g., discrete Lagrangian dynamics), would be a good method for the application of QC to fluid dynamics. 

\begin{figure}[htpb!]
\centering{
\includegraphics[scale=0.25]{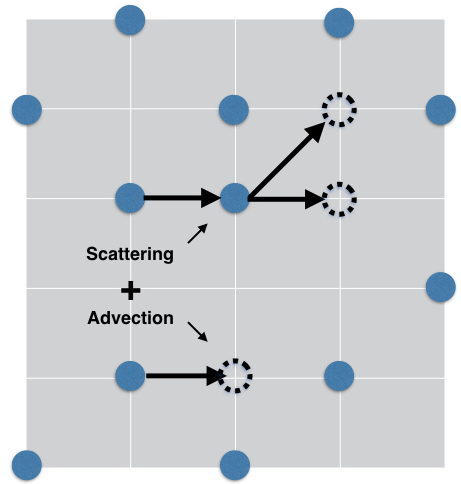}
}
\caption{Schematic of an LBM simulation}
\end{figure}

\subsection{LQC 1: Quantum Lattice Gas Automaton and Phase Coherent Quantum Networks}
This method provides insights into \textit{one} of the physical frameworks for generating a map from classical fluid dynamics to QC. As an aside, though this method has been proposed quite a while ago, only theoretical and classical computer simulations of the QC method had been done (due to the absence of a real QC at the time), so the implementation on presently available QCs remains to be established. The key idea here is derived from Quantum Lattice Gas Automaton (QLGA) \cite{meyer1996quantum,meyer2002quantum}, which is a quantum extension of the classical lattice gas system. 

As a simple illustration let us consider a 1D lattice system. The classical lattice gas tags every particle with instantaneous positions and velocities $(\mathbf{x,v})_{i}$, where the velocities $\mathbf{v}_{i}$ at every lattice site points either to the left or the right. With this scheme, we generate an ensemble of 1D state configurations, which evolves according to a local evolution map. This mapping, like the LBM in Equation 7, is a combination of advection and scattering processes. 

Now its quantum counterpart, the QLGA, prepares quantum superpositions of the classical states. For a single particle 1D lattice of length N, this results in each site having 2 pseudo occupation slots (q), corresponding to the left (l) and right (r) streaming particles with associated probabilities. This means that we now have a 2 qubit system $|\psi \rangle_{i} = \alpha|q_{1} q_{2}\rangle_{i} + \beta|q_{1} q_{2}\rangle _{i} $ and $q_{1}, q_{2} \in \{l,r\}$ sitting on each site, hopping to adjacent sites with a basis set $\{|lr\rangle, |ll\rangle, |rr\rangle, |rl\rangle \}$. The scattering processes of these qubits is given by the scattering matrix $\hat{S}$ that captures the interactions, while the advection is given by $\hat{A}$. If L and R are left and right scattering probability amplitudes (i.e., the probability that a particle travelling left as it enters a site continues leftwards, etc), the scattering matrices for the cases of one and two qubits (for some p) would be given by
\begin{equation}
 S_{1} =   \begin{pmatrix}
L & R \\
R & L
\end{pmatrix} = \begin{pmatrix}
\cos p & i\sin p\\
i\sin p & \cos p
\end{pmatrix} 
\end{equation}

\begin{equation}
 S_{2} =   \begin{pmatrix}
1 & 0 & 0 & 0 \\
0 & L & R & 0 \\
0 & R & L & 0 \\
0 & 0 & 0 & \theta
\end{pmatrix} = \begin{pmatrix}
1 & 0 & 0 & 0 \\
0 & \cos p & i\sin p & 0 \\
0 & i\sin p & \cos p & 0 \\
0 & 0 & 0 & \theta 

\end{pmatrix} ,
\end{equation}

\noindent where $|L|^2 + |R|^2 = 1 = |\theta|^2$ and $\theta$ represents the relevant multi-particle scattering events of the delta-function type. Thus the dynamics of the advection and scattering steps could be summarised quantitatively as follows, with the propagation of the left and right travelling wavefunctions:
    \begin{align}
        \Psi(\mathbf{r}, t+1 ; \rightarrow) &= L\Psi(\mathbf{r}-1, t ; \rightarrow) + R\Psi(\mathbf{r}+1, t ; \leftarrow) \\
        \Psi(\mathbf{r}, t+1 ; \leftarrow) &= L\Psi(\mathbf{r}+1, t ; \leftarrow) + R\Psi(\mathbf{r}-1, t ; \rightarrow).
    \end{align}
Here, L and R scale as p, and the evolution of such a process is unitary and preserves the norm. The sum of these two wavefunctions satisfies the Schr\"odinger equation. Interestingly we can recover both the Dirac and the Schr\"odinger equation with limits $\delta \mathbf{r} \rightarrow 0$ and $\delta \mathbf{r}^2 \rightarrow 0$ as $\delta t \rightarrow 0$, which is done using the standard Chapman-Enskog asymptotic closure \cite{meyer1996quantum,boghosian1997quantum,boghosian1998quantum}. 

A fact worth mentioning is that the quantum effect of the scattering operator causes a local entanglement in a specific lattice zone radius, while the advection operator acts globally, causing superposition of configuration states as well as a global entanglement. The QLGA setup was originally used in this form to perform quantum many body system simulations \cite{boghosian1997quantum,boghosian1998quantum,boghosian1998simulating,lloyd1996universal,abrams1997simulation}, and was later modified into what is known as the Phase Coherent Quantum Lattice Network to compute mesoscopic and macroscopic fluid dynamics, and to study quantum mechanically the diffusion equation and Burgers flow \cite{yepez1998lattice,yepez2001quantum,yepez2001quantuma,yepez2002quantum}. 

These amended versions naturally evolved in response to the need for minimizing impediments such as errors due to large entanglement and noises that depend on the environment. The main difference in these extensions is that, instead of the qubits representing state superpositions, they directly replace the classical bits that store site information in classical LBM. For instance, in a model with 4 qubits per site, with each lattice site having 4 nearest neighbors, one can encode $2^4$ complex numbers per site. Thus, every site now acts as a small 4 qubit QC and many such QCs can be connected via a lattice network to form a coherent quantum network. In general, for an $N$-site lattice with $N$\textsubscript{q} qubits per lattice (also the number of nearest neighbors), we have a lattice QC made of a total of $N$\textsubscript{T} = $N*N$\textsubscript{q} qubits. We would thus have a $2^{N_{T}}$-dimensional Hilbert space composed of $2^{N_{q}}$ dimensional submanifolds corresponding to local site-specific Hilbert spaces and, with $\mathbf{C}$ representing the coefficient matrix and $q_{i}$'s being the qubits on every site, the total wavefunction would be  
 \begin{align}
        \Psi(\mathbf{r_{1}},.. ,\mathbf{r_{N}},t) & = \sum_{\psi(\mathbf{r_{i}},t)} \mathbf{C} \times (\psi(\mathbf{r_{1}},t) \otimes .... \otimes \psi(\mathbf{r_{N}},t)) \\
         & = \sum_{\psi_{q_{i}}} \tilde{\mathbf{C}} \times (\psi_{q_{1}} \otimes .... \otimes \psi_{q_{N_{T}}}).
    \end{align}

The evolution operator for this lattice gas simulation is obtained by integrating the corresponding Schr\"{o}dinger equation and, as explained above, this unitary evolution operator would now correspond to the product of the unitary advection ($\hat{A}$) and scattering ($\hat{S}$) matrices, giving us 
\begin{equation}
    \Psi(\mathbf{r},t+\Delta t) = e^{i\hat{\mathbf{H}}\Delta t/\hbar} \Psi(\mathbf{r},t) = \hat{A}\hat{S}\Psi(\mathbf{r},t).
\end{equation}
 
We might now ask which of the mesoscopic fluid dynamical observables can be computed. For this, let us take a Bravais Lattice with lattice site positions given by $\mathbf{R}$, while the corresponding unit vectors $\mathbf{r}$ and wavefunctions are given by the propagating Bloch vectors with a lattice-specific periodicity (i.e., we can reach any site from any other site by advancing through an integral multiple of the lattice periodicity, yielding a total of steps to be $k_{i} \leq N*(period)$). Recasting a lesson from the classical lattice gas, we can now compute the occupancy probability, as well as mass and momentum densities \cite{yepez2001quantum} as follows. 

(a) The occupancy probability for the quantum case is straightforward and is just the average of the number operator $\hat{\mathbf{n}}_{\mathbf{R}} = \hat{c}_{\mathbf{R}}^{\dag} \hat{c}_{\mathbf{R}}$. This can be measured on many practical QCs like NMR-QC (nuclear magnetic resonance) by Quantum State Metrology (which we shall describe in the sections to follow). From the basic postulate of quantum mechanics, the average is given by the trace of the operator taken with the lattice's density matrix, thus giving the occupation probability to be
\begin{equation}
    P_{\mathbf{R},t} = \text{Tr}[(|\Psi(\mathbf{R},t)\rangle \langle \Psi(\mathbf{R},t)|) \hat{\mathbf{n}}_{\mathbf{R}}] =  \text{Tr}[\mathbf{\rho}(t) \hat{\mathbf{n}}_{\mathbf{R}}].  
\end{equation}

(b) If $d$ is the lattice spacing, we can directly write down the mass and momentum densities ($\xi$ and $\xi \mathbf{v}$) as
\begin{align}
        \xi(\mathbf{r},t) & = \lim_{d \rightarrow 0} \sum_{k_{i},\mathbf{R_{1}}}^{\mathbf{R_{N}}} m \text{Tr}[\mathbf{\rho}(t) \hat{\mathbf{n}}_{{\mathbf{R}}_{k_{i}}}],  \\
        \xi(\mathbf{r},t)\mathbf{v}(\mathbf{r},t) & = \lim_{d \rightarrow 0} \sum_{k_{i},\mathbf{R_{1}}}^{\mathbf{R_{N}}} m v^{2} \mathbf{r}_{(\mathbf{R}mod N_{q})} \text{Tr}[\mathbf{\rho}(t) \hat{\mathbf{n}}_{{\mathbf{R}}_{k_{i}}}].
    \end{align}
Now, as the limit reaches the continuum (i.e., with higher lattice resolution), we could use quantum observables such as the number operator to estimate mesoscopic quantities. Finally, one can also write the mesoscopic transport equation as 
\begin{equation}
     P_{\mathbf{R}+d\mathbf{r},t+\Delta t}  =  P_{\mathbf{R},t} + \langle \Psi(\mathbf{R},t)|\hat{S}^{\dag}\hat{\mathbf{n}}_{\mathbf{R}}\hat{S} - \hat{\mathbf{n}}_{\mathbf{R}}|\Psi(\mathbf{R},t)\rangle,
\end{equation}
 since \\
 $$
  \langle \Psi(\mathbf{R},t)|\hat{S}^{\dag}\hat{\mathbf{n}}_{\mathbf{R}}\hat{A}^{\dag}|\Psi(\mathbf{R},t+\Delta t)\rangle  =  \langle \Psi(\mathbf{R},t)|\hat{S}^{\dag}\hat{\mathbf{n}}_{\mathbf{R}}\hat{A}|\Psi(\mathbf{R},t)\rangle.
 $$

These equations and operators can be simulated by appropriately recasting them in terms of generalized $U_{3}$ gates (IBMQ) given by 3 parametric unitary gates, which are the Euler angles \cite{yepez2001quantum,IBM}:
 \begin{equation}
 U_{3} =   \begin{pmatrix}
\cos(\frac{\theta}{2}) & e^{-i\lambda}\sin(\frac{\theta}{2}) \\
e^{i\phi}\sin(\frac{\theta}{2}) & e^{i(\phi+\lambda)}\cos(\frac{\theta}{2}) 
\end{pmatrix}.
\end{equation}
Here $0 \leq (\phi, \lambda) < 2\pi$ and $0 \leq \theta \leq \pi$. Thus, the takeaways of this method are the following: (1) it provides an understanding of the translation of classical LBM calculations to its quantum analog, thus enabling newer methods for building QC circuits to simulate fluid dynamics; (2) it gives an idea of how one could make use of quantum lattice properties and entanglement to map it to macroscopic properties of the flow; and (3) though a clear estimate on scaling behavior, compared to classical LBM, is an open avenue, it seems obvious that one can gain over classical LBM, in both space and time complexity of the problem, since we are using state space configurations in terms of quantum superpositions and are performing simultaneous evolutions of these states. We shall also examine a more recent variant of this method that is more amenable to implementation and expected to utilize exponential speedup due to better way of quantum superposition.

\subsection{LQC 2: Dirac Equation and the Pseudo Spin Boson system}

\begin{figure*}[htpb!]
\centering{
\includegraphics[scale=0.72]{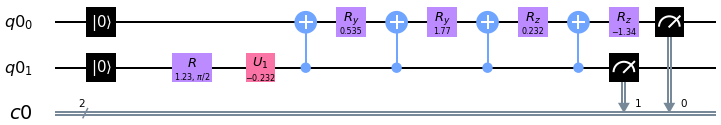}}
\caption{Circuit for \textit{amplitude} loading 4 complex numbers}
\end{figure*}
A variant procedure is the construction of a map from the classical LBM to the Dirac equation \cite{succi1991lattice,benzi1992lattice,mezzacapo2015quantum}. We shall not dwell on details of this method but provide only an outline of the fluid dynamical aspects. This method generates a map, using what is known as a coupled pseudo-spin bosonic system, which is amenable for implementation on a trapped ion QC or a superconducting QC. We first note that we can translate Equation (7) into a ``Majorana type Dirac equation" of the form \cite{fillion2013formal}
\begin{equation}
    i\frac{\partial \mathbf{\Psi}}{\partial t} + i\hat{\mathbf{A}}_{pb}\nabla_{pb} \mathbf{\Psi} = \mathbf{M} \mathbf{\Psi},
\end{equation}
where $\hat{\mathbf{A}}$ is the quantum analog of the advection matrix (a Clifford operator---since it is given by Pauli matrices), while $\mathbf{M}$ is the quantum representation of the mass term, which is Hermitian. Since we know that Clifford operators (isomorphic in $\mathbb{R}^3$ with Pauli operators) do not simultaneously commute, we will need to have diagonal advection operators and symmetric and imaginary scattering matrices. The idea is essentially to map the mass density to a corresponding probability density of a wavefunction that is embedded in an appropriate ``Fock space" of the given bosonic modes. For instance, in the 1D case, we would have a single bosonic wavefunction distribution as $|\Psi(x,t)\rangle _{i} = \int dx \text{P}(x)|x\rangle _{i}$, where P(x) holds the information of the mass distribution. (For a higher dimensional Fock space with 2 bosonic modes in a 2D lattice, $|\Psi(\mathbf{x},t)\rangle _{i} = \int dx_{1}dx_{2} \text{P}(x_{1},x_{2})|x_{1}\rangle|x_{2}\rangle$.) For producing different such distributions, we need to have an external parametrized knob that can control them. This task is accomplished by what is known as the ``pseudo spins", which are coupled to the bosonic modes as $|\Psi\rangle = \sum_{i}P_{i}|s\rangle_{i} \otimes |\Psi(\mathbf{x},t)\rangle$. The operators to diagonalize the advection matrix would, in the second quantized Dirac picture, be the following \cite{mezzacapo2015quantum}:
\begin{equation}
  e^{-i\hat{A}\Delta t/ \hbar} = exp\Big(-\frac{i\Delta t}{\hbar}\frac{\pi}{2\sqrt{2}} \big(  \text{Z}_{1}\otimes I_{2} +  \text{X}_{1} \otimes \sigma_{2}^{pb} \big) \Big).
\end{equation}
Here $\text{Z}_{i}$ and $\text{X}_{i}$ are the usual Pauli operators on i\textsuperscript{th} mode and $\sigma$\textsuperscript{pb} refers to the pseudo spin. On the other hand, the scattering operator $\hat{S}$, which is essentially a non-unitary type evolution step, is made ``pseudo-unitary" by making its evolution dependent on an ancillary qubit, which acts as the control (refer to \cite{mezzacapo2015quantum} for detailed methodology); finally, this $\hat{S}$ is decomposed into a weighted sum of two unitary operators as $e^{-i\hat{S}\Delta t/ \hbar} = e^{\hat{U}_{1}+\alpha \hat{U}_{2}}$. Successive application of these operators on the initial lattice wavefunction can be done by a standard and useful trick of decomposing the evolution operators via a suitable Lie-Trotter-Suzuki decomposition, which allows one to translate such operations as a quantum circuit using generalised $U_{3}$-type gates. Importantly, resources needed to do this operation is a polynomial in the degrees of freedom, while being sub-polynomial in error. Finally, we add suitable quantum metrology to extract the final state. The entire process has been illustrated for a simple advection-diffusion equation in \cite{mezzacapo2015quantum}. Also, since this method, unlike previous ones, replaces classical bits directly by qubits, the pseudo-spin system that is used manifests quantum superposition, thus allowing one to exploit the exponential speed up of the superposition principle of QC; it would certainly be interesting to validate this expectation on present QCs. 
 
\subsection{LQC 3: Quantum to Classical Mapping}

This method has often been used in early theoretical calculations by performing a mapping from a d-dimensional partition function for quantum systems to a (d+1)-dimensional partition function for classical systems. In condensed matter systems, this is accomplished by mapping classical to quantum lattices by means of classical Monte-Carlo methods \cite{sondhi1997continuous,hsieh2016d}; in conformal field theories, this is achieved by understanding gravitational bulk-boundary correspondence \cite{aharony2000large,polyakov1987gauge}, etc. Though we will not describe details, we believe that this has the potential for hybridizing either LQC 1 or LQC2 along with quantum-to-classical mapping for solving fluid dynamical problems. 

In this method, one basically computes a quantum partition function $\text{Tr}(e^{-H\beta})$ from the Feynman imaginary time path integral \begin{equation}
    Z = \sum_{r,r_{i}}\langle r|e^{-H\Delta \beta /N }|r_{1}\rangle \langle r_{1}| ... |r_{N}\rangle \langle r_{N}|e^{-H\Delta \beta /N }|r\rangle,
\end{equation} which is exactly the classical (d+1)-dimensional partition function, except that the extra dimension is replaced by space instead of time; it can be solved using the conventional transfer matrix type calculation. Now if one sets up a lattice-style fluid dynamics problem in 2D, it would be equivalent to solving a 1D quantum lattice problem with this map. (There also have been efforts to map d dimensions $\rightarrow $ d dimensions for purposes of understanding classical statistical mechanics in quantum mechanical terms \cite{somma2007quantum}.) This is now being actively applied to connect classically simulated annealing to quantum annealing methods to solve optimization problems. In fact, we shall discuss them separately to see how machines based on quantum annealing are being used to solve fluid dynamics problems.% Put scaling argument of David Meyer. Proc Royal Soc.

\section{CONTINUUM SIMULATIONS}
Simulations done in a true continuum sense (i.e., no lattice models) ultimately boil down to preparing a mix of computational tools or algorithms that can emulate standard mathematical methods in quantum mechanics. To this end, we draw the reader's attention to some currently implementable quantum algorithms and circuits that one could use in fluid dynamics.

When we refer to an implementable quantum algorithm, we mean quantum circuits that can be constructed from known coherent set of quantum logic gates and measurement processes. A complete computational process or simulation of a fluid dynamic problem involves three essential steps: (1) Initial data input or loading; (2) Processing and generating new data; and (3) Reading the processed information to obtain results. Each of these steps, though obvious, involves nontrivial operations in QC. We will now take a closer look at them now.
 %fig of data loading 3 steps 

\subsection{Data loading}

The data inputs could be user-defined computational parameters or initial conditions of an ODE/PDE integrator, etc. Classically we input and store data, for instance in C++, by writing algorithmically \textbf{int a = 10;}, so that \textbf{a} holds the value 10. At the machine level, the data are  ``written" as a magnetic inscription of local magnetic polarities on a hard disk. We now ask how we could do the same exercise on a QC, i.e., store the value 10 in \textbf{a}. The storing of such a qubit at machine level comes in a variety of ways mentioned earlier. Here, we shall not explain how these physical realisations work, but dwell more on the algorithmic level, considering the machine level as an existent oracle. We can load a classical bit of information onto a quantum computer in two ways: \\
 
\noindent(1) \underline{Amplitude loading}: This method, also known as state preparation or the initialization method \cite{plesch2011quantum,nielsen2002quantum}, where one initializes a specific qubit state with the user-defined complex probability amplitudes of quantum superposition. That is, the input data are loaded in the form of complex probability amplitudes of wavefunctions. One of the algorithms widely followed on practical QCs (like IBMQ) is outlined in Algorithm 1 \cite{shende2006synthesis}, though one can always come up with custom circuits to construct a given state.
In this method, called the recursive quantum multiplexor algorithm, one starts with the required state and designs a circuit to transform the required state to all 0s; thus, the inverse circuit prepares our required state starting from all 0 states. As a demonstration, suppose we want to load four complex values such as $(i)$, $(2+i)$, $\sqrt{2}i$, $1$ (in this order). Since we can store the N values in $\log_{2}N$ qubits, we need a 2 qubit system that looks like the following:
\begin{equation}
    |\Psi\rangle = \frac{1}{3}\big( \mathbf{i} |00\rangle + \mathbf{(2+i)} |01\rangle + \mathbf{\sqrt{\mathbf{2}}i} |10\rangle + \mathbf{1} |11\rangle \big) .
\end{equation}
The required set of unitary operations, obtained by following Algorithm 1 \cite{shende2006synthesis}, is represented in the quantum circuit shown in Figure 8. Thus the probabilities of these states are essentially the magnitudes of these states (i.e $\frac{1}{9}, \frac{5}{9}, \frac{2}{9} \text{ and } \frac{1}{9}$). Upon performing state tomography (see later), the final state when computed with Qiskit---the IBM Quantum Experience platform---is shown in Figure 9, which almost exactly matches the required state. 

\begin{algorithm}
\caption{Amplitude Loading: Quantum State Preparation}
\SetKwInput{KwInput}{Input}                % Set the Input
\SetKwInput{KwOutput}{Output}              % set the Output
\DontPrintSemicolon
  
  \KwInput{$c_{1},c_{2} \cdots c_{n}$}
  \KwOutput{ $U_{load}$ , $|\Psi\rangle_{final} = c_{1}|00...0\rangle + c_{2}|00...1\rangle + ... + c_{n}|11...1\rangle $}
  %\KwData{Testing set $x$}

% Set Function Names
  \SetKwFunction{amp}{AMPLITUDELOAD}
  \SetKwFunction{disent}{DISENTANGLE}
  %\SetKwFunction{FSub}{Sub}
 
% Write Function with word ``Function''
  \SetKwProg{Fn}{Function}{:}{}
  \Fn{\amp{$c_{1},c_{2} \cdots c_{n}$}}{
        $U = \mathbb{I}$\;
        $|\Psi_{final}\rangle = |00..0\rangle$ (On Quantum Device)\;
        $|\Psi\rangle_{temp} \gets c_{1}|00...0\rangle + c_{2}|00...1\rangle + ... + c_{n}|11...1\rangle$ (On Classical Device)\;
        \While{( $|\Psi\rangle_{temp} == |00....0\rangle $)}{$|\Psi\rangle_{temp}$ = \disent{$|\Psi\rangle_{temp}$} Least Significant Bit\;
        Choose $\alpha_{i}$ , $\beta_{i}$ and global phase $\phi$, such that:\; \While{($|q_{i}\rangle \gets c_{i}e^{i\phi}|0\rangle$)}{
        $U_{z} = \mathbf{R_{z}(\beta_{i})}|q_{i}\rangle$\;
        $U_{y} =\mathbf{R_{y}(\alpha_{i})}|q_{i}\rangle$\;
        }
         
        $(U_{zy})_{i} \gets U_{y}\otimes U_{z} $\;
        $U \gets (U_{zy})_{i}\otimes U $\;
        
        }

        Now $U = (U_{zy})_{n}\otimes \cdots \otimes (U_{zy})_{1}$\;
        $U_{load} = U^{\dag}$\;
        $|\Psi_{final}\rangle = U_{load}|\Psi_{final}\rangle = c_{1}|00...0\rangle + c_{2}|00...1\rangle + ... + c_{n}|11...1\rangle $\;
        \KwRet $U_{load}$, $|\Psi_{final}\rangle$\;
  }

% Write Function with word ``Def''
  %\SetKwProg{Fn}{Def}{:}{}
  %\Fn{\FSub{$first$, $second$}}{
   %     a = first\;
    %    b = second\;
        %sum = first - second\;
     %   \KwRet sum\;
  %}
  %\;

  \SetKwProg{Fn}{Function}{:}{\KwRet}
  \Fn{\disent{$|\Psi\rangle_{temp}$}}{
       $|\Psi\rangle_{local} = |\Psi\rangle_{temp} \gets c_{1}|00...q_{1}\rangle + c_{2}|00...q_{2}\rangle + ... + c_{n}|11...q_{n}\rangle $\;
        %b = 10\;
        %Sum(5, 10)\;
        %Sub(5, 10)\;
       % print Sum, Sub\;
        \KwRet $|\Psi\rangle_{local} \gets c_{1}|00...\rangle \otimes|q_{1}\rangle + c_{2}|00...\rangle \otimes|q_{2}\rangle + ... + c_{n}|11...\rangle \otimes|q_{n}\rangle$\;
  }
\end{algorithm}
\begin{figure}[htpb!]
\centering{
\includegraphics[scale=0.52]{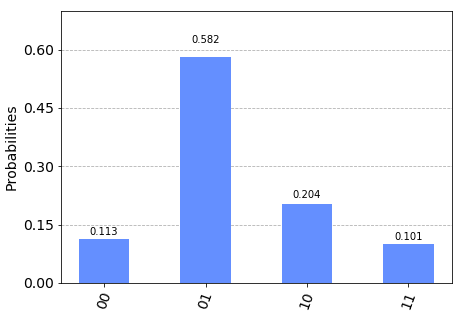}
}
\caption{Required state prepared on IBMQ}
\end{figure}

\begin{figure}[htpb!]
\centering{
\includegraphics[width=7cm,height=7cm]{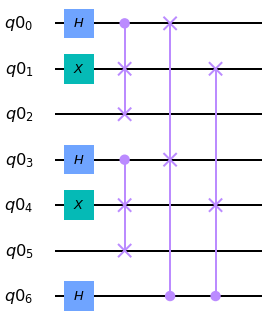}
}
\caption{Circuit for \textit{state} loading 4 classical bits}
\end{figure}

\begin{figure*}[htpb!]
\centering{
\includegraphics[scale=0.5]{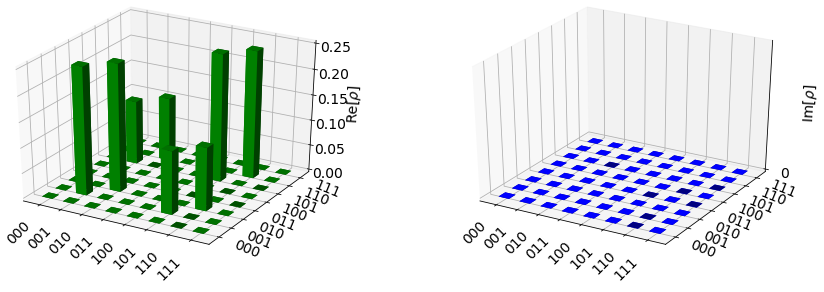}
}
\caption{The density matrix of the data loaded 3 qubit state}
\end{figure*} 
%\newpage
\noindent(2) \underline{State loading}: Instead of using complex amplitudes, one may store the data directly as the state ket vectors. Suppose we want to load the decimal number 10, whose binary form is 1010. By state loading, we mean that there is some state that looks like $|\psi\rangle = |\phi\rangle \otimes (|1\rangle + |0\rangle + |1\rangle + |0\rangle) $. To do this, since we need 4 basis states, one has to construct a state such as $ |\psi\rangle = (|0\rangle|0\rangle \otimes|1\rangle + |0\rangle|1\rangle\otimes|0\rangle + |1\rangle|0\rangle\otimes|1\rangle + |1\rangle|1\rangle\otimes|0\rangle) $, where we see that the third qubit holds the value we need, 1010. To prepare such a state, we use a combination of Controlled-SWAP, X, H and Toffoli gates as shown in Figure 10. (A detailed explanation of such circuit designs is found at \cite{cortese2018loading}.) 

Figure 11 shows the density matrix $\rho = \sum_{i} p_{i}|\psi\rangle_{i}\langle\psi|_{i} $ of the prepared state. Looking at the third bit in Figure 11, which reads 1,0,1,0 corresponding to the highest peaks, we have the demonstration that the algorithm has encoded the data into our qubits. Though one can see 6 qubits in the circuit, 4 of them are actually ancillary, i.e., they are dummy qubits needed only for the processing and can be discarded at the end of the circuit by disentangling them from our data. Therefore, as the number of bits (N) grows larger, it can be shown that the most optimal circuit would need $\log_{2}(N)$ qubits. This scaling in data loading is necessary if we have to build quantum processing circuits with exponential scaling. 
 
\subsection{Output Measurements}

After loading and producing new data, the data are processed by a series of unitary operations required by the problem and the final state of the qubits is the output result we seek. The challenge is to efficiently estimate these final states, which are the complex probability amplitudes and the state vectors. This process is called quantum state estimation or quantum state metrology or quantum state tomography \cite{nielsen2002quantum,vogel1989determination,leonhardt1997measuring,nawrocki2015introduction}. 

This process is not straightforward because, in order to probe these qubits, one would have to perform a measurement, which physically means interacting with the wavefunctions. Such interactions, apart from inducing noise, also collapse the wavefunctions to one of the basis states resulting in smudging and loss of data and results. The most elementary way to evaluate a state is to perform a von Neumann projective measurement on the state along (say) the z-axis that forms the eigenvector of the computational basis states $|0\rangle$ and $|1\rangle$. But, quantum mechanics tells us that projective measurements project the state to a particular basis vector and may not always depict the complete information represented by a wavefunction. To get a total estimate of a given superposition state, one would have to do something better. 

The next simplest method is the empirical probability estimation by ensemble averaging, i.e., conducting several identical experiments to generate a set of outputs and perform projective measurements every time, to collect the probability statistics of each state under superposition (whose accuracy obviously increases with the ensemble size). In order to make our discussion clear, let us briefly discuss the meaning of quantum measurements.

QM Postulate: \textit{Measurement is an operator A that acts on a set of quantum states $\psi_{i}$ to yield a physical observable eigenvalue ``a" with the following properties:}
\begin{enumerate}
    \item The probability of obtaining outcome ``a" is
    \begin{equation}
        P(a) = \langle \psi_{i}| A_{a}^{\dag} A_{a} |\psi_{i} \rangle.
    \end{equation}
    \item The post measurement state would be
    \begin{equation}
        |\psi_{f} \rangle = \frac{A_{a} |\psi_{i} \rangle}{\sqrt{\langle \psi_{i}| A_{a}^{\dag} A_{a} |\psi_{i} \rangle}}~. 
    \end{equation}
\end{enumerate}
Now, these measurements give us only the probability of a given eigenvalue ``a'' from a given state $|\psi\rangle$. But this output is ''weak" in the sense that it is only from one single state and the measurement process, and other successive measurements may not yield the same result even though we start out from identically prepared initial states, because of noise and decoherence. Thus we might have to average over many such experiments and measurements. Since output of each experiment differs, it naturally creates an ensemble of states \{$|\psi_{i}\rangle$\}. 

As an aside, we note that this fact could be used advantageously for turbulence simulations. Since even slightly different initial conditions lead to different dynamics, due to the inherent chaotic nature of the system, in general, the present ensemble automatically represents an ensemble of many turbulent evolutions. The standard way that quantum mechanics suggests for characterizing such an ensemble of states, whose exact form we have to probe, is to use the density operator formalism. If $p_{\alpha}$ is the probability of obtaining a state \{$|\psi_{\alpha}\rangle$\}, the density operator is given by $\rho = \sum_{\alpha}p_{\alpha}|\psi_{\alpha}\rangle \langle \psi_{\alpha}|$. We may restate the previous measurement postulate in terms of $\rho$ as
\begin{enumerate}
    \item The probability of obtaining outcome ``a" is
    \begin{equation}
        P(a) =  \text{Tr}(A_{a}^{\dag} A_{a}\rho). 
    \end{equation}
    \item Its post measurement state would be
    \begin{equation}
        \rho_{f} = \frac{A_{a}\rho A_{a}^{\dag}}{\sqrt{\text{Tr}(A_{a}^{\dag} A_{a}\rho)}}. 
    \end{equation}
\end{enumerate}

The density operator has the property that $\text{Tr}(\rho) = 1$ i.e., probabilities (non-negative eigenvalues) sum up to unity. Since we are only interested in obtaining the statistics of each superposition state, we ask the question: What type of measurement procedure respects the positivity and completeness property of the density operator as well as yield the probability amplitudes of the states in the ensemble? The answer is termed POVM Measurements (Positive Operator Valued Measure). The POVM elements constitute a set of operators \{$P_{VM}^{a}$\} $\equiv$ \{$A_{a}^{\dag}A_{a}$\} that are constrained to be positive, since $\langle \psi_{i}| P_{VM}^{a} |\psi_{i} \rangle = \langle \psi_{i}| A_{a}^{\dag} A_{a} |\psi_{i} \rangle = P(a) \geq 0$, and complete, i.e., $\sum_{a}P_{VM}^{a} = \sum_{a}A_{a}^{\dag} A_{a} = \mathbb{I}$. In fact, as one can easily observe, we can even get the measurement operator corresponding to a given POVM element by $\sqrt{P_{VM}^{a}} = A_{a}$. 

We are generally not interested in the measurement itself but only in the statistics. Thus the POVM provides a clean way of doing this without worrying about the state itself. Another important advantage of positive definiteness of these operators is that, for a given set of non-orthogonal states, a POVM set of operators $\{P_{VM}^{a}, P_{VM}^{b} \text{ and } (P_{VM}^{c} = \mathbb{I}-(P_{VM}^{a}+P_{VM}^{b}))\}$ can be used to compute their statistics or distinguish the states. Thus, the output of sandwiching the operators between state vectors gives us the statistics. The operators are designed to create a unique one-to-one mapping from the operator space to a particular output state. So any positive output, resulting from the first two operators in our set, points uniquely to a specific state, while an output from the third operator implies that we cannot comment anything about the state. So, essentially such a measurement process never lets us go wrong in identifying states, but at the cost of being unable to comment about the output from one of its operators. With this machinery we are ready to look at quantum state tomography.

\subsection{Quantum State Tomography} 

Let us begin with a density matrix representing an unknown quantum state that needs to profiled. Say we have just one set of non-orthogonal qubit states. Experimentally it is impossible to construct the quantum state from just one copy of the states. But we can make several POVM measurements from multiple copies and compute the statistics. The set of operators \{$\mathbb{I}/2$, X/2 , Y/2 , Z/2 \} that form a set of orthonormal operators can be used to expand the density matrix as
\begin{equation}
    \rho = \frac{1}{2}\Big(\text{Tr}(\rho)\mathbb{I} + \text{Tr}(X\rho)X + \text{Tr}(Y\rho)Y + \text{Tr}(Z\rho)Z\Big).
\end{equation}
The expectation of these operators can be obtained by $\text{Tr}(X\rho)$. Now each of these expectation values can be estimated by repeated measurements. Once we have a large sample size with good estimates of each of these operator outputs, we can reconstruct the density operator of the unknown state. The standard deviation of the estimate is $1/\sqrt{N}$, where N is the ensemble size, as one would expect for a Gaussian random variable in the large $N$ limit. This in essence is the picture of Quantum State Tomography. The same procedure can be extended to multiple qubit systems as well. This procedure is usually achieved in practice by a few popular techniques \cite{coles2018quantum} such as
\begin{enumerate}
    \item \textit{Simple Inversion}
    \item \textit{Regression Fits}
    \item \textit{Maximum Likelihood Estimates}
    \item \textit{Bayesian Methods.}
\end{enumerate}
\begin{table*}[htpb!]
\centering
\caption{\textbf{Quantum Algorithms}} 
\vspace{4mm}
\begin{tabular}{|l|cc|r|}
\hline
{\#}& {{Algorithm}} & {Description/Used in} & {Complexity/Speed-up}\\
\hline
%&&  &  \\
%&\textbf{Algorithm}& \textbf{Description/Used in}& \\
%&&  &  \\
1&Quantum Teleportation \&  & Inter-circuit data communication \&  & - \\

&Entanglement \cite{nielsen2002quantum}&  a fundamental block of many algorithms  &  \\
&&  &  \\

2&Superdense Coding \cite{nielsen2002quantum} & Data compression \& communication &  Compression Ratio 2:1 \\
&&  &  \\

3&Quantum Fourier Transform & DFT, Phase Estimation, Period Finding, &\textbf{Q}: [$\Theta(n^2),\Theta(n\log n)$]  \\ 
&(QFT) \cite{nielsen2002quantum} & Arithmetic, Discrete log \& spectral methods  & \textbf{C}: $\Theta(n2^{n}$) \small{(n=\#gates)} \\

&&  &  \\

4&Quantum Phase Estimation \cite{nielsen2002quantum} & Quantum phases, Order Finding, Shor's  &\textit{O}(t\textsuperscript{2}) operations**\\
&& Algorithm, HHL, Amplitude Amplification  & $t=n+\left \lceil{\log \Big(2 + \frac{1}{2\epsilon}\Big)}\right \rceil $ \\
&& \& Quantum Counting \cite{brassard1998quantum} &  \\
&&  &  \\

5&Grover's Search \cite{nielsen2002quantum,grover1997quantum} \& & Data search, Amplitude Estimation, Function & \textbf{Q}: \textit{O}($\sqrt{N}$)\\
&  Amplitude Amplification \cite{brassard2002quantum} &  minima, approx. \& Quantum counting   &\textbf{C}: \textit{O}(N) \small{(N=\#ops)}\\
&&  &  \\

6& Matrix Product Verification \cite{Buhrman} & Verifies AB=C? (n$\times$n matrices)&\textbf{Q}: $\leq$\textit{O}($n^{5/3}$)); \textbf{C}: \textit{O}($n^2$)\\
&&  &  \\

7& Quantum Simulation \cite{lloyd1996universal,nielsen2002quantum} & Integrates Schr\"odinger equation, HHL, & \textit{superpoly}\\
&&All Hamiltonian system simulations   \Big($e^{-iHt/\hbar}$\Big)  &\textit{poly}(n,t): n=dof, t= time \\
&&  &  \\

8& Gradients \cite{jordan2005fast,zoo} & Computes gradients, convex optimisation & \textit{quadratic - superpoly}  \\
&&   volume estimation, minimising quadratic forms  &  \\

&&  &  \\

9& Partition Function \cite{zoo} \&  & Evaluate/approx partition functions& \textit{quadratic - superpoly} \\
&Sampling & Pott's, Ising Models \& Gibbs sampling & \\

&&  &  \\

10& Linear Systems \& & Solves A\textbf{X}=\textbf{b} for eigen values \& vectors&\textit{superpoly - exponential}  \\
&HHL Algorithm \cite{harrow2009quantum,zoo} & ODEs, PDEs, simultaneous eqns. & \\
&&  Optimisation, Finite Element Methods etc  & \\

&&  &  \\

11& ODE \cite{berry2014high,zoo}  & Integrates $\dot{\mathbf{x}}=\alpha(t)(\mathbf{x})+\beta(t)$ \& similar forms &\textit{superpoly - exponential} \\
&&  &  \\

12& Wave Equation \cite{costa2019quantum} & Integrates $\ddot{\phi}=c^2 \nabla ^2 \phi$ \& similar forms  &\textit{superpoly - exponential} \\
&&  &  \\

13& PDE / Poisson Equation \cite{cao2013quantum,arrazola2019quantum,zoo} & Integrates $ - \nabla ^2 \phi(\mathbf{x}) = b(\mathbf{x}) $ and & \textit{superpoly - exponential}  \\
&& PDEs of similar forms: $\mathbf{D}\phi(\mathbf{x})=b(\mathbf{x})$ & \\
&&  &  \\

14& QFT Arithmetic \cite{ruiz2017quantum} & QFT based: + , - , * , mean , weighted sum& \textit{superpoly - exponential} \\
&&  &  \\

15& Function Evaluation \cite{hadfield2018quantum} & (Ex) inverse, exponentiation.. etc & varies \\
&& for State Loaded data  & \\
&&  &  \\

16& VQE and QAOA \cite{IBM} & Computes optimisation type problems & varies\\
&&  &  \\

17& Quantum Annealing \cite{DWave} & Computes optimization type problems & varies\\

\hline
\end{tabular}
\tablenotes{Q/C-Quantum/Classical; VQE-Variational Quantum Eigen solver; QAOA-Quantum Approximate Optimisation Algorithm; \\ ** t = number of qubits with n-bit phase approximation and $\epsilon$ error.  }
Specific references to the above algorithms can be found at \cite{zoo,IBM,coles2018quantum,montanaro2016quantum}
 \end{table*}
Simple illustrations and explanations of a few of these techniques are given in \cite{coles2018quantum}. More advanced quantum metrology techniques, which also make use of quantum phase estimation methods that could be used for fluid dynamics applications, are discussed in \cite{xu2018turbulent}. As an example of a simple ensemble averaged measurement, we demonstrate tomography of two entangled qubits. For this, we prepare a sample entangled Bell state
\begin{equation}
|\psi\rangle = \frac{1}{\sqrt{2}}(|00\rangle + |11\rangle) .
\end{equation} via Amplitude Loading (as shown in Figure 12) and try to estimate its probability amplitudes by running on the IBMQ. As one can clearly see, each qubit has the probability 1/2, which is exactly what we want to estimate experimentally. As one can see the results from Figure 13, the histogram peaks properly with almost equal probabilities ($\approx 0.5$) at $|00\rangle$ and $|11\rangle$. The small but finite probabilities of the other two states is due to quantum errors and decoherence in the system. With this, we conclude our brief discussion on output measurements. 
\begin{figure}[htpb!]
\centering{
\includegraphics[scale=0.7]{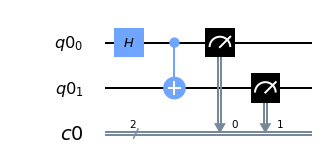}
}
\caption{The quantum circuit for entanglement preparation and measurement}
\end{figure}
\begin{figure}[htpb!]
\centering{
\includegraphics[scale=0.5]{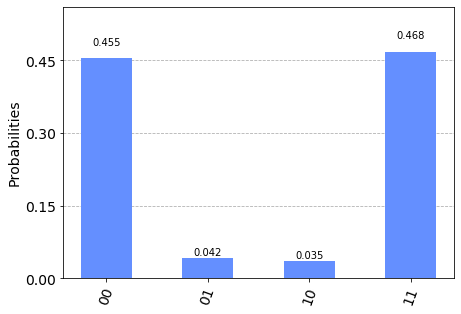}
}
\caption{Probability amplitude estimates}
\end{figure}
 
\subsection{Data Processing}

Having briefly discussed inputs and outputs of a quantum computational process, we shall now examine the quantum algorithms that one needs to use to generate, process and manipulate data. A comprehensive and updated collection and descriptions of most of the available quantum algorithms can be found in \cite{zoo,coles2018quantum,nielsen2002quantum,IBM,montanaro2016quantum}. To keep our discussion contained, we choose only a few important ones that are possible candidates for fluid dynamics simulations in Table 2. For the example problem of a 1D steady Stokes flow (creeping flow),
\begin{align}
    &\nabla^2 u(x)  =  \nabla p(x) \\
    &\nabla \cdot u(x)  =  0,
\end{align}

\noindent a sample working procedure could look like this:\\ 

\noindent \underline{Step A - Data Loading}: First, we need to initialise $u(x)$ and the right hand side with an appropriate numerical initial iterate value. This is done by the quantum state preparation as described earlier. \\

\noindent \underline{STEP B - Data Acquisition}: To integrate the equation, one could think of several numerical methods to do this:
    \begin{enumerate}
        \item Finite Element Method: Like any FEM, we discretize the system first. This dicretization procedure, along with boundary conditions, yields a matrix which performs the differential operation. In our example, the first equation is a Poisson equation. We could use any variant of Algorithm \#13 to start solving this problem. In general, solving such an FEM setup boils down to solving a matrix inversion problem, which is done by Algorithm \#10 (HHL) =  Alg \#3 + Alg \#4 + Alg \#7. Based on which algorithm we choose, we can get up to an exponential speed up in computation. This is demonstrated in \cite{cao2013quantum}.
        \item Pseudospectral Methods: We can use Algorithm \#3 (QFT) to first map both the LHS and RHS to the spectral space, first computing the derivatives in the spectral space and then using HHL to invert. This method could also yield exponential speed up.
        \item Amplitude amplification: Further, we can append Algorithm \# 5 to perform amplitude amplification to amplify the probability of obtaining the right answer in every experimental run.
        \newline
    \end{enumerate}
\noindent \underline{STEP C - Output Measurements}: Once we obtain the eigenvalues and eigenvectors, we can perform a quantum state tomography to extract the results and store them in classical registers.

A key observation is that, based on the choice of numerical integration, we can achieve up to an exponential speed up. This has great potential for simulating the Navier-Stokes equations. For instance, the currently available pseudospectral DNS codes face a major bottleneck with the FFTW steps that need to be mapped out for computing derivatives. Even if we could set up a hybrid classical-quantum integration, where only the FFTW step are replaced by the QFT, we could achieve exponential speed up. To motivate this direction, a QFT demonstration is given in the following section.

\subsection{Quantum Fourier Transform}
The Quantum Fourier Transform \cite{nielsen2002quantum} is very similar to the Discrete Fourier Transform performed by the currently available FFTW routines. To keep the discussion concrete, consider the simple example of discretizing the domain and sampling the function at (say) four points. Now, the DFT of this function $f$ at these four points may be written as 
\begin{equation}
    F[k] = \sum_{j=0}^{j=N-1} f[j] exp\Big(2\pi i\frac{jk}{N}\Big) .
\end{equation}
Here, the amplitudes $f[j] = \Big\{ f[0],f[1],f[2],f[3]$\Big\} are being Fourier transformed to $F[k] = \Big\{F[0],F[1],F[2],F[3]$\Big\}. Very similarly, the Quantum Fourier Transform (QFT) is a unitary operation that transforms as 
\begin{equation}
   \sum_{i=0}^{i=N-1} \alpha_{i}|i\rangle \mapsto \sum_{i=0}^{i=N-1} \beta_{i}|i\rangle ,
\end{equation}
where
\begin{equation}
    \beta_{k} = \frac{1}{\sqrt{N}}\sum_{j=0}^{j=N-1}\alpha_{j}exp\Big(2\pi i\frac{jk}{N}\Big).
\end{equation}
Equivalently, if we set $\omega^{jk}= exp\Big(2\pi i\frac{jk}{N}\Big)$
\begin{equation}
    |\alpha \rangle \mapsto \frac{1}{\sqrt{N}}\sum_{\beta=0}^{\beta=N-1} \omega^{jk}|\beta\rangle .
\end{equation}
Note that the states in the computational basis $|0\rangle, |1\rangle, |2\rangle, |3\rangle $ can be expressed in the binary form as well, which is $|00\rangle, |01\rangle, |10\rangle, |11\rangle $. With this step, we obtain the transformations of the sampled amplitudes to be
\begin{align*}
    \beta_{0} &= \frac{1}{2}[\alpha_{0} + \alpha_{1} + \alpha_{2} + \alpha_{3}]\\
    \beta_{1} &= \frac{1}{2}[\alpha_{0} + i\alpha_{1} - \alpha_{2} - i\alpha_{3}]\\
    \beta_{2} &= \frac{1}{2}[\alpha_{0} - \alpha_{1} + \alpha_{2} - \alpha_{3}]\\
    \beta_{3} &= \frac{1}{2}[\alpha_{0} - i\alpha_{1} - \alpha_{2} + i\alpha_{3}],
\end{align*}
caused by the unitary transformation operator $U_{QFT}$. If $\omega = e^{i\pi /2}$ the operator is given by
\begin{equation}
    U_{QFT} = \frac{1}{2} \begin{pmatrix}
1 & 1 & 1 & 1\\
1 & i & -1 & -i\\
1 & -1 & 1 & -1\\
1 & -i & -1 & i\\
\end{pmatrix}
=
\frac{1}{2} \begin{pmatrix}
1 & 1 & 1 & 1\\
1 & \omega & \omega^2 & \omega^3\\
1 & \omega^2 & 1 & \omega^2\\
1 & \omega^3 & \omega^2 & \omega\\
\end{pmatrix}.
\end{equation}
The quantum circuit that implements such a 2-qubit QFT is shown in Figure 14 below.
\begin{figure}[htpb!]
\centering{
\includegraphics[scale=0.7]{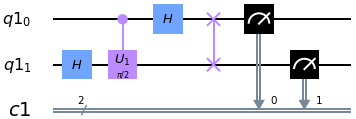}
}
\caption{The quantum circuit for 2-qubit QFT}
\end{figure}
To illustrate its action, let us prepare the state already shown in Figure 9 using Amplitude Loading and run the QFT on it. The expected outputs of the Fourier transform for these amplitudes, computed analytically, is $\beta_{0} = 0.574 ,\beta_{1} = 0.037, \beta_{2} = 0.306 ,\beta_{3} = 0.138 $, while the results obtained from the QFT are shown in Figure 15.
\begin{figure}[htpb!]
\centering{
\includegraphics[scale=0.5]{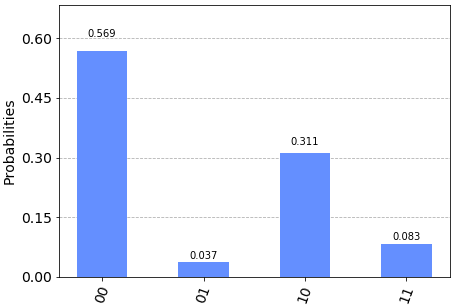}
}
\caption{The QFT output}
\end{figure}
The QFT results are $\beta_{0} = 0.569 ,\beta_{1} = 0.037 ,\beta_{2} = 0.311 ,\beta_{3} = 0.083$. The precision can improve with better Quantum State Tomography, while the computation of QFT is exponentially faster than DFT.

\subsection{Quantum Turbulence}

It would obviously be instructive to look at quantum fluid dynamics as well. A quantum fluid, like any other quantum system, is described by its corresponding many-body interacting Hamiltonian and its evolution is governed by a corresponding Schr\"odinger equation (a quick insight can be obtained from \cite{barenghi2014introduction,barenghi2014experimental}. The evolution described by this equation represents an analytically consistent way of obtaining a proper microscopic evolution. Since tracking $\approx 10^{23}$ particles is computationally impractical, given that macroscopic observables are what we wish to understand, there is a huge motivation for developing macroscopic equations of motion. The complete many-body quantum simulation may well be possible with a powerful quantum computer, but one may also be able to develop more efficient versions of numerically integrating different model equations. Both methods provide insight on the right descriptions of quantum turbulence and vortex reconnection. The following QC tasks are possible candidates for studying quantum turbulence.

\subsubsection{Quantum simulation}

The most microscopic description is that of the Hamiltonian picture. For instance, consider a Bose-Einstein condensate Hamiltonian and its corresponding Schr\"odinger equation:
    \begin{equation}
        \mathcal{H}_{BEC} = -\frac{\hbar^2}{2m}\mathbf{\nabla}^2 + V_{ext}(\mathbf{r}) + U_{int}(\mathbf{r-r'}), 
    \end{equation}
       \begin{align}
        \frac{\partial}{\partial t}|\Psi_{BEC}\rangle &= \frac{1}{i\hbar}\Big[-\frac{\hbar^2}{2m}\mathbf{\nabla}^2 + V_{ext}(\mathbf{r}) + U_{int}(\mathbf{r-r'})\Big]|\Psi_{BEC}\rangle \\
        |\Psi_{BEC}\rangle &= e^{-i\mathcal{H}_{BEC}t/\hbar}|\Psi_{BEC0}\rangle.
    \end{align}
Now this calls for a many-body simulation. With the growing surge of QC methods of many-body algorithms, we could perform a quantum simulation (Algorithm \#7) \cite{nielsen2002quantum} to evolve this equation. Among the methods being developed, the Trotter decomposition method (or the Lie-Trotter-Suzuki decomposition) is worth mentioning. If P and Q are Hermitian operators (that need not commute), we have, for any t
    \begin{equation}
        \lim_{N\rightarrow\infty}(e^{iPt/N}e^{iQt/N})^N = e^{i(P+Q)t} .
    \end{equation}
    Now the same result can be used to derive higher order corrective equations such as:
    \begin{multline}
        e^{i(P+Q)\delta t} = (e^{iP\delta t}e^{iQ\delta t}) + \mathcal{O}(\delta t^2)\\ = (e^{iP\delta t/2}e^{iQ\delta t}e^{iP\delta t/2}) + \mathcal{O}(\delta t^3) .
    \end{multline}
    Now this is a very useful result, since we can take our $\mathcal{H}_{BEC}$ and split it into $\mathcal{H}_{BEC}=\sum_{i=1}^{n}\mathcal{H}_{i}$ Hamiltonian operators acting on smaller sub-systems spanning a local Hilbert space. If $[\mathcal{H}_{i},\mathcal{H}_{j}] = 0 \forall i,j$ and $\hbar = 1$, we have
    \begin{equation}
        e^{-i\mathcal{H}_{BEC}t} = e^{-i\mathcal{H}_{1}t}e^{-i\mathcal{H}_{2}t}...e^{-i\mathcal{H}_{n}t} .
    \end{equation}
    But if commutation is not imposed, a similar correction is obtained:
    \begin{equation}
        e^{-i\mathcal{H}_{BEC}t} = e^{-i\mathcal{H}_{1}t/2}e^{-i\mathcal{H}_{2}t}e^{-i\mathcal{H}_{1}t/2} + \mathcal{O}(\delta t^3) .
    \end{equation}
Following this procedure, each time-step operator can be decomposed into basic unitary logic gates and a corresponding evolution circuit can be constructed. The following model equations would be amenable to numerical integration using QC algorithms. 

\subsubsection{The two-fluid model}

Simulating the Landau's equations would be useful for those looking at quantum fluids at low velocities and with no quantum vortices, since this model works best for irrotational and incompressible flows. The idea would be to build on the previously discussed algorithms for dealing with ODEs and PDEs procedures to integrate the following equations:
    \begin{align}
        \frac{\partial \mathbf{u_{1}}}{\partial t} + \mathbf{u_{1}}\cdot \nabla \mathbf{u_{1}} &= -\nabla\Big(\frac{p_{1}}{\rho_{1}}\Big),\\
        \frac{\partial \mathbf{u_{2}}}{\partial t} + \mathbf{u_{2}}\cdot \nabla \mathbf{u_{2}} &= -\nabla\Big(\frac{p_{2}}{\rho_{2}}\Big) + \frac{\nu}{\rho_{2}} \nabla^2 \mathbf{u_{2}} .
    \end{align}
Here the subscripts 1 and 2 correspond to superfluid and normal fluid, respectively. Let us now look at methods that includes quantum vortices as well.

\subsubsection{Gross-Pitaevskii model}

Along with a few approximations and assumptions, we can use the standard trick of Madelung Transformation to establish a relationship between the BEC wavefunction and fluid macroscopic properties such as density and velocity. This is the Gross-Pitaevskii equation
    \begin{multline}
        \frac{\partial}{\partial t}|\Psi_{cond}\rangle = \frac{1}{i\hbar}\Big[-\frac{\hbar^2}{2m}\mathbf{\nabla}^2  + V_{ext}(\mathbf{r}) +\\ + U_{int}|\langle\Psi_{cond}|\Psi_{cond}\rangle|\Big]|\Psi_{cond}\rangle .
    \end{multline}
The built-in assumptions are: (a) Though the actual BEC wavefunction is a sum of the actual condensate wavefunction and the pertubative term, at $T \sim 0$, we say $\Psi_{BEC} \approx \Psi_{cond}$. (b) Length scales are of the order of the vortex cores. (c) Only contact interactions are allowed $U_{int} = U\delta(\mathbf{r-r'})$. This model is the nearest microscopic description, yet has many limitations. A detailed outlook could be obtained from \cite{barenghi2014introduction,barenghi2014experimental}.

\subsubsection{Vortex filament model}

The next level would be to move to scales greater than the vortex core sizes. We visualise the fluid as a ensemble of arcs of quantum vortices and track these vortex arcs $\mathbf{l}$, which is the vortex filament model. The evolution of these arcs is given by
    \begin{equation}
        \frac{d\mathbf{l}}{dt} = \mathbf{u_{sa}} + \mathbf{u_{f}} ,
    \end{equation}
where $\mathbf{u_{sa}}$ is the self-advecting velocity of the vortex and $\mathbf{u_{f}}$ is the mutual friction between the normal fluid arc surface. The computationally heavy step to be done by the QC is the evaluation of the Biot-Savart integral 
    \begin{equation}
        \mathbf{u_{i}} = \frac{\Gamma}{4\pi}\oint \frac{(\mathbf{l}-\mathbf{l'})}{|\mathbf{l}-\mathbf{l'}|^3} \times d\mathbf{l} ,
    \end{equation}
to compute $\mathbf{u_{sa}}$, $\Gamma$ being the circulation of the vortex filaments.
\subsubsection{HVBK model}

This model, obtained from a slight amendment of Landau's equation, gives the best description for the largest scales, much larger than the core size. The additional terms are the mutual friction force $\mathbf{f_{mf}}$ and the arc tension force $\mathbf{f_{T}}$. This model too has limitations owing to its assumptions on the vortex arc orientations. The equations are (let us call $\mathbf{F}= \mathbf{f_{mf}} + \mathbf{f_{T}}$ :
    \begin{align}
        \frac{\partial \mathbf{u_{1}}}{\partial t} + \mathbf{u_{1}}\cdot \nabla \mathbf{u_{1}} &= -\nabla\Big(\frac{p_{1}}{\rho_{1}}\Big) - \mathbf{F} , \\
        \frac{\partial \mathbf{u_{2}}}{\partial t} + \mathbf{u_{2}}\cdot \nabla \mathbf{u_{2}} &= -\nabla\Big(\frac{p_{2}}{\rho_{2}}\Big) + \frac{\nu}{\rho_{2}} \nabla^2 \mathbf{u_{2}} + \frac{\rho_{1}}{\rho_{2}}\mathbf{F} .
    \end{align}

\section{VARIATIONAL SOLVERS AND QUANTUM ANNEALERS}

The last method to be outlined is based on variational optimization. Suppose we want to solve the conventional CFD problem of simulating a Stokes flow by using a discretization solver such as Gauss-Seidel or Jacobi. The problem reduces to solving for eigenvalues of the form $A\mathbf{x}=B$ using the HHL algorithm. It can also be solved as an optimization problem. That is, we define a cost function such as the difference between the LHS and RHS of the eigenvalue problem, and iterate and modify $\mathbf{x}$ so as to minimise the cost function to 0. Classical methods include algorithms such as gradient descent, steepest descent, conjugate gradient method, etc. Such optimization procedures could be used for QC as well. \\

  \noindent \underline{A. Variational Quantum Eigen (VQE) Solver}. The idea stems from the principle of quantum mechanics for solving the eigenvalue problems variationally. It is usually done as a hybrid of quantum and classical computing. So far, VQE has been applied for different condensed matter and quantum chemistry problems, but it can be extended to other problems as well. On a hybrid machine, the steps are noted below. Detailed descriptions can be found in \cite{IBM,coles2018quantum,peruzzo2014variational}.
  
    \begin{enumerate}
        \item Consider a matrix P with one of its eigenvectors $|\psi_{p}\rangle$. Then we know that the $|\psi_{p}\rangle$ is invariant in the sense of $P|\psi_{p}\rangle=p|\psi_{p}\rangle$, where $p$ is the corresponding eigenvalue. Let us regard P as a Hamiltonian, which is a positive definite Hermitian matrix, with positive and real eigenvalues. Thus, the expectation value of the Hamiltonian is $\langle\psi|\mathcal{H}|\psi\rangle \geq 0$. The smallest eigenvalue $p_{min}\leq\langle\psi|\mathcal{H}|\psi\rangle$ corresponds to the ground-state energy of the system($\geq0$), which can be estimated by Algorithm 2.

        %\item Begin with a guess to the initial state $|\psi(\mathbf{k})\rangle$. The initial state is prepare by acting a ``well" parametrised unitary operator such as the U3($\mathbf{k}$) gate(IBMQ).  
        %\item Now, we perform the expectation value evaluation $\langle\psi|\mathcal{H}|\psi\rangle$ on the QPU and record the values.
        %\item Based on the output of QPU, feed this to a QPU along with the parameters $\mathbf{k}$ and perform an optimisation search using a classical algorithm to update the parametrised guess $|\psi(\mathbf{k})\rangle$.
        %\item Then feed this updated $|\psi(\mathbf{k})\rangle$ to the QPU again.
        %\item Iterate until convergence.
        \item Thus while a QPU computes expectation values, the CPU runs an optimisation algorithm; together they can be used to estimate the eigenvalue and ground state configurations. 
    \end{enumerate}

\noindent \underline{B. Quantum Approximate Optimisation Algorithms (QAOA)}: Generally, combinatorial optimization methods may not be tractable with polynomial resources. Other than developing problem specific methods, approximate algorithms such as QAOA can be handy. The goal is to take a discrete variable as an input, which could be strings of binaries such as $\mathbf{x} = x_{1}...x_{n}$, where $\mathbf{x}_{i} \in \{0,1\}$ defines a cost function E($\mathbf{x}$) that needs to be maximized. The cost function is essentially a map from $E(\mathbf{x}): \{0,1\}^{n} \mapsto \mathbb{R}$. The QAOA \cite{farhi2014quantum,coles2018quantum,IBM} thus forms the set of algorithms which does exactly this, and guarantees that the approximation ratio $\alpha$ satisfies
    \begin{equation}
        \alpha=\frac{E(\mathbf{x})}{E_{max}} \geq \alpha_{opt} .
    \end{equation}

\begin{algorithm}[]
\caption{Variational Quantum Eigenvalue Solver}
\SetKwInput{KwInput}{Input}                % Set the Input
\SetKwInput{KwOutput}{Output}              % set the Output
\DontPrintSemicolon
  
  \KwInput{$\mathcal{H},|\psi(\mathbf{k})\rangle,\mathbf{k} (\text{parameter}), \epsilon (\text{tolerance}) $}
  \KwOutput{$ |\psi(\mathbf{k})\rangle_{opt}$,$E_{opt} $}
  %\KwData{Testing set $x$}

% Set Function Names
  \SetKwFunction{vqe}{VQE}
  \SetKwFunction{optim}{CLASSICAL\_OPTIMISER}
  \SetKwFunction{init}{INITIALISE}
  %\SetKwFunction{FSub}{Sub}
 
% Write Function with word ``Function''
  \SetKwProg{Fn}{Function}{:}{}
  \Fn{\vqe{$\mathcal{H},|\psi(\mathbf{k})\rangle,\mathbf{k}, \epsilon$}}{
        \init{$|\psi(\mathbf{k})\rangle$}\;
        (Can also use $U_{3}(\mathbf{k})$ gates (Eqn. 19) for parametrisation)\;
        \While{( $p_{min} \leq \epsilon \text{ }\&\&\text{ } P | \psi(\mathbf{k})\rangle==p_{min}|\psi(\mathbf{k})\rangle \approx p|\psi_{p}\rangle)$)}{
        $E_{opt} = \langle\psi|\mathcal{H}|\psi\rangle$ (ground state energy estimate)\;
        $|\psi(\mathbf{k})\rangle_{opt} = | \psi(\mathbf{k})\rangle = $ \optim{$| \psi(\mathbf{k})\rangle$}
        }
        
        \KwRet $ |\psi(\mathbf{k})\rangle_{opt}$,$E_{opt} $\;
  }

  \SetKwProg{Fn}{Function}{:}{\KwRet}
  \Fn{\optim{$| \psi(\mathbf{k})\rangle$}}{
       Optimise $| \psi(\mathbf{k})\rangle$ based on parameter $\mathbf{k}$ on a CPU\;
       \KwRet $ |\psi(\mathbf{k})\rangle_{opt}$
  }
\end{algorithm}

\noindent \underline{C. Quantum Annealing}: This method is now widely used to run on what are essentially known as Quantum Annealer Machines (which in essence are not quantum computers) such as those produced by companies such as DWave. The physical principle here is to use the quantum analogue of simulated annealing that one uses to solve optimization problems in classical physics, but the phenomena of quantum tunnelling sets the quantum version apart from the classical one. This phenomena is exploited for scanning fast through different minima of a given energy landscape of a cost function. In the classical Monte Carlo, one would have to thermally excite the system to jump the energy barrier to the next minimum, while in the quantum case, even with tall energy barriers and with a certain thin barrier width, one can ''tunnel" to the adjacent minimum as shown schematically in Figure 16. 

The DWave system does exactly this. With a combination of quantum tunnelling, quantum entanglement and a transverse field bias, it can perform quantum annealing efficiently \cite{battaglia2005optimization,DWave} and find optimal solutions via minimization. This has already been put to use to study the Navier-Stokes channel flow \cite{ray2019towards}. In this work, one first converts the NS equation into a discretized version and sets up the problem as an $A\mathbf{x}=B$ eigenvalue problem as usual. Later, this is numerically investigated by converting the problem into an optimization type setup called the Quantum Unconstrained Binary Optimisation (QUBO) supported by the DWave machine.
\begin{figure}[htpb!]
\centering{
\includegraphics[width=8cm,height=6cm]{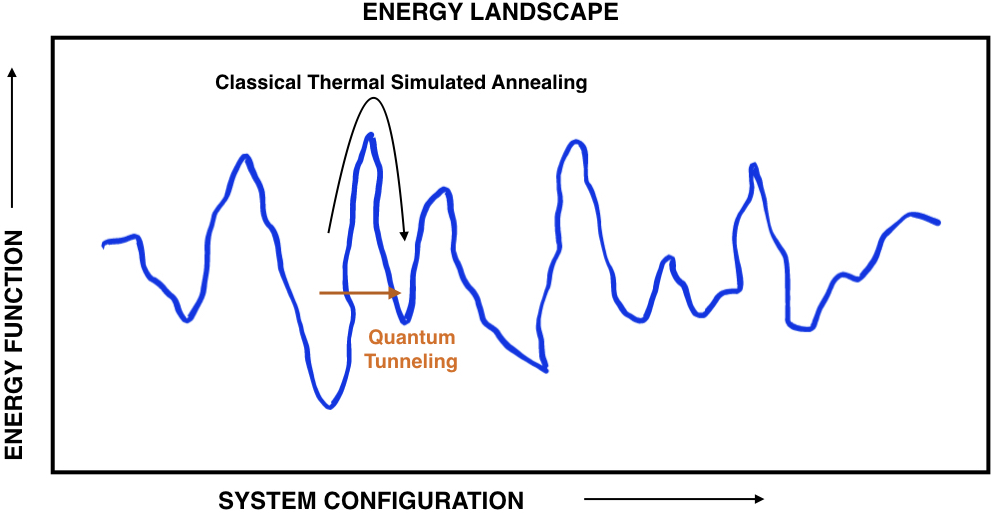}
}
\caption{Quantum Annealing}
\end{figure}

These variational and optimization methods are already being used for commercial applications such as traffic flow management and finance management with very big quantum annealers such as DWave which is offering 5000 qubits. Though the complexity estimates of these methods varies and is not yet clearly established, it still offers a lucrative quantum protocol that can solve optimization problems with a decent speedup. 

\begin{figure*}[htpb!]
\centering\includegraphics[height=8cm,width=12cm]{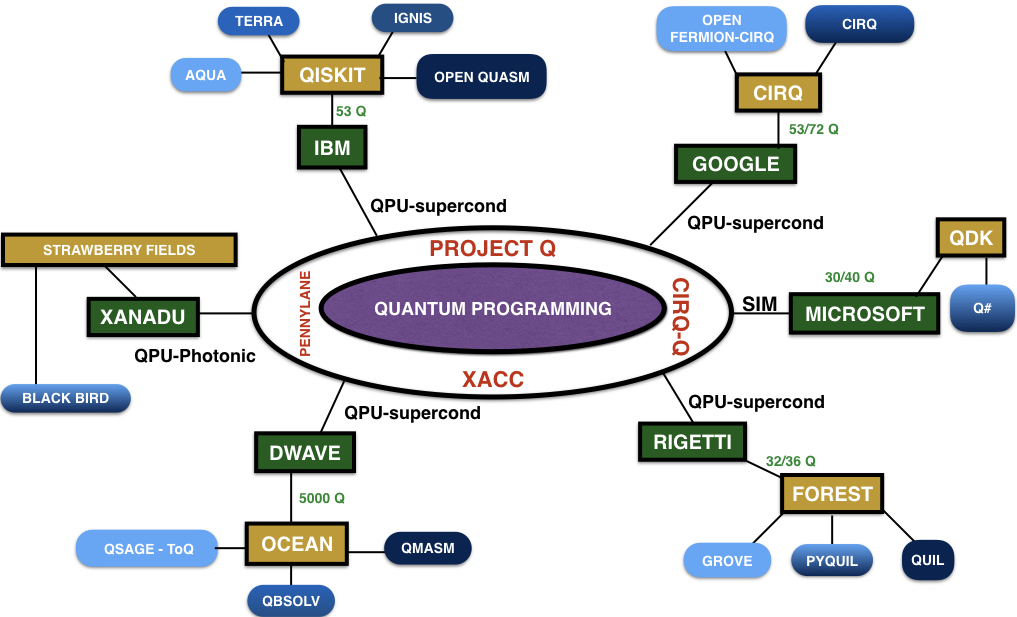}
\end{figure*}
\begin{figure*}[htpb!]
\centering\includegraphics[scale=0.3]{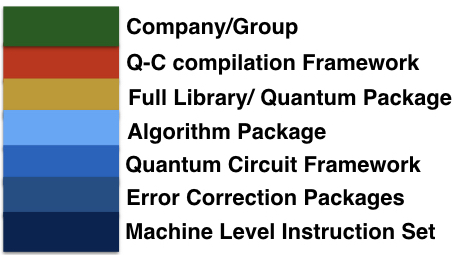}
\caption{Different QCs and Architectures}
\end{figure*}

\section{QUANTUM PROGRAMMING AND MACHINES}

Coming to the implementation and the actual execution of these ideas and algorithms, what we need are (a) efficient quantum computer simulators to test the correctness of quantum algorithms, and (b) real quantum computing devices to execute and assess the quantum advantage of practical QCs. There are now a large and growing number of efforts that have already built, and are trying to build, better quantum computers. Each of these QCs is being implemented using different quantum physical realizations and quantum materials that would be robust against external noise and decoherence, called Quantum Processing Units (QPU) or Quantum Processors. Given a QPU, the process of programming a set of quantum algorithms and converting them into forms understandable by a quantum machine, using instructions of suitable programming languages, is called quantum programming.

The different programming languages, though seemingly similar, vary in terms of instruction sets and the actual physics governing the operation of the QC. Different quantum programming methods and packages are being developed by different QC companies. Concentrating on computational fluid dynamics, different workstations ranging from simple PCs to massive supercomputers have been used. Similarly, the currently available quantum devices and programming packages are summarized in Figure 17. Each of these available quantum programming kits have their own strengths and weaknesses; for instance, certain devices are better set for optimization type problems compared to others. As mentioned earlier, all the demonstrations shown so far were done using the Qiskit programming kit of IBMQ based on a transmon type superconducting qubit. In fact, we are currently in an era analagous to when classical computing had computers and storage devices with capacities ranging from a few bits to a few bytes; we now have qubits instead of cbits. Though the number of qubits is small, one should note that computing capacity can be exponentially larger compared to its classical counterparts.

Finally, let us consider the IBMQ 54-Qubit machine for some specific remarks.
\begin{enumerate}
    \item QFT: In most DNS simulations, the FFT step is highly time consuming. With 54 Qubits, we can encode and compute the FFT of $2^{54} \approx 10^{16}$ complex numbers exponentially faster than the classical version. This by itself, if implemented coherently, can greatly speed-up DNS calculations.
    \item DNS Grid Sizes: With a 54-Qubit machine we can store and compute on
    \begin{enumerate}
        \item 1D:  $\leq 10^{16} $ meshes
        \item 2D:  $\leq 10^{8}\times10^{8} $ meshes
        \item 3D:  $\leq 10^{5}\times10^{5}\times10^{5} $ meshes
    \end{enumerate}
Here, each of these mesh sizes is far higher than the largest available DNS computations at present.
\end{enumerate}

With these quantum algorithmic subroutines, at this stage itself, exponential speedups are possible, and their robust implementation can ease the computational challenges facing DNS.
%\section{A Summary of Computational Complexity}

%%Use table environment for a table in one column

%\begin{table}[htb]
%\caption{Table fitting in a single column}\label{tableExample}
%\begin{tabular}{|l|cccc|r|}
%\hline
%one& two &three&four&five&six\\
%1&2&3&4&5&6\\
%aaa&bbbb& ccccc&dddd&eeeee&ffffff\\
%\hline
%\end{tabular}
%%use \tablenotes{footnote} to get the table foot note
%\tablenotes{Sample table footnote}
%\end{table}

%%Use table* environment to get the table spanning both the columns

%\begin{table*}[htb]
%\caption{Caption text here}\label{secondTable}
%\begin{tabular}{|l|cccccccc|r|}
%\hline
%\textbf{head1}&\multicolumn{8}{|c|}{\textbf{head2}}&\textbf{head3}\\
%\hline
%one& two &three&four&five&six&seven&eight&nine&ten\\
%1&2&3&4&5&6&7&8&9&10\\
%aaa&bbbb&cccc&ddddd&eee&ffff&ggggg&hhhhhhhh&iiii&jjjjjj\\
%\hline
%\end{tabular}
%\tablenotes{table footnote here}
%*table spanning both the columns
%\end{table*}

%%An example of a figure

%\begin{figure}[!t]
%\centering{
%\includegraphics[width=.8\columnwidth]{fig1.eps}
%}
%\caption{caption goes here}\label{figOne}
%\end{figure}

%%An example of a double column figure
%%Use figure* environment

%\begin{figure*}
%\centering\includegraphics[height=.15\textheight]{fig2.eps}
%\caption{caption spanning two columns}
%\centering\includegraphics[height=.25\textheight]{fig3.eps}
%\caption{caption here}
%\end{figure*}

\section{CONCLUSIONS}
We have discussed and illustrated a selected collection of quantum methods and tools for QCFD simulations. Most of these tools present at least a quadratic speedup, sometimes superpolynomial or exponential, compared to their classical counterparts. This in itself is an incentive for the QCFD study. The existence of commercially available QCs such as IBMQ provides an additional thrust to this entire effort. We wish to bring to the reader's attention that quantum error correction and decoherence reduction methods form a key field of study, whole scope is to reduce noise-related errors and make QC codes more robust and accurate. For fluid dynamicists, the onset of QC provides a unique and exciting opportunity to study the subject in a completely new way. Progress calls for familiarity with this new paradigm of computing, building and putting together newer and existing quantum algorithms for QCFD solvers. Since we are still in the early stage, it would be wise to perform hybrid computations, where some functions are done on a QPU with the others on a CPU or GPU. As a concluding thought, this paper is intended to motivate the pursuit of new directions of computational fluid mechanics that have the potential of a huge impact.

%%Appendix

%\appendix

%\section{An appendix section}
%Text goes here.
%\begin{equation}
%x=a+b+c
%\end{equation}
%\section{Another appendix section}
%Text goes here.
%\begin{equation}
%y^2=ax+b+c
%\end{equation}
%%Use section* for acknowledgements

\section*{Acknowledgement}
We acknowledge the use of the IBM Q and IBM Q Experience platform for this
work. The views expressed are those of the authors and do 
not reflect the official policy or position of IBM or the IBM
Q team. We thank J\"org Schumacher, Dhawal Buaria and Kartik P Iyer for insightful discussions.

%%use \balance somewhere in the left column of the last page to balance the two columns in the end page

%%References section

\end{document}